\documentclass[final,3p,11pt]{elsarticle_no_date}

\usepackage{amssymb}

\def\pr{{\partial}}
\def\eps{{\epsilon}}
\def\bx{{\bf x}}
\def\bv{{\bf v}}
\def\be{{\bf e}}

\begin{document}

\begin{frontmatter}

\title{Steady viscous flows in an annulus between two
cylinders produced by vibrations of the inner cylinder}
\author{K. Ilin\corref{cor1}}
\ead{ki502@york.ac.uk}
\author{M. A. Sadiq}
\ead{msa502@york.ac.uk}
\cortext[cor1]{Corresponding author}
\address{Department of Mathematics, University of York,
Heslington, York, YO10 5DD, U.K.}

\begin{abstract}

\noindent
We study the steady streaming between two infinitely long circular
cylinders produced by small-amplitude transverse vibrations
of the inner cylinder about the axis of the outer cylinder. The Vishik-Lyusternik
method is employed to construct an asymptotic expansion of the solution of
the Navier-Stokes equations in the limit of
high-frequency vibrations for Reynolds numbers of order of unity. The effect of the Stokes drift of fluid particles
is also studied. It is shown that it is nonzero not only within the boundary layers but also in higher order
terms of the expansion of the averaged outer flow.

\end{abstract}


\end{frontmatter}

\setcounter{equation}{0}
\renewcommand{\theequation}{1.\arabic{equation}}

\section{Introduction}
\label{Sec1}

\noindent
In this paper, we study two-dimensional oscillating flows of a
viscous incompressible fluid between two infinitely long circular
cylinders. The outer cylinder is fixed, and the inner
cylinder performs small-amplitude harmonic oscillations about the
axis of the outer cylinder. It is well-known that high-frequency
oscillations of the boundary of a domain occupied by a viscous
fluid not only produce an oscillating flow, but can also lead to
appearance of a steady flow, which is usually called the steady
streaming (see, e.g., \cite{Riley2001}). Here we are interested in this steady part of the flow.
The basic parameters in
our study are the inverse Strouhal number $\alpha$ and the
dimensionless viscosity $\nu$ (the inverse Reynolds number), defined as
\begin{equation}
\alpha=\frac{V_{0}^{*}}{a\omega}, \quad
\nu=\frac{1}{Re}=\frac{\nu^{*}}{V_{0}^{*}a} \label{1.1}
\end{equation}
where $V_{0}^{*}$ is the amplitude of the velocity of the
oscillating cylinder, $a$ is its radius, $\omega$ is the angular
frequency of the oscillations, and $\nu^{*}$ is the kinematic
viscosity of the fluid. Parameter $\alpha$ measures the ratio of the amplitude of the
displacement of the oscillating cylinder to its radius and is
assumed to be small: $\alpha \ll 1$.
One more dimensionless parameter which is widely used in literature is the `streaming Reynolds number'
$Re_{s}=V_{0}^{*2}/\omega\nu^{*}=\alpha/\nu$,
which plays the role of the Reynolds number for the steady part of the flow.
Note that for small $\alpha$, $Re_{s}=O(1)$ corresponds to $\nu=O(\alpha)$,
i.e. to high Reynolds numbers ($Re=1/\nu \gg 1$).

\vskip 3mm \noindent
The steady streaming produced by an
oscillating cylinder both in an unbounded fluid and in a
cylindrical container had been studied by many researchers. A
review of early works can be found in \cite{Riley1967}. Regular perturbation
analysis had been used to study both an
oscillating cylinder in an unbounded fluid and the case of the
flow between two cylinders in
\cite{Holtsmark1954, Skavlem1955} for the flow regimes with $\nu
\gg 1$ ($Re \ll 1$). In \cite{Wang1968}, the
method of matched asymptotic expansions had been used to construct
an asymptotic solution for an oscillating cylinder in an unbounded
fluid for $\nu =O(1)$ ($Re =O(1)$). In \cite{Bertelsen1973}, the
theoretical results of \cite{Holtsmark1954, Skavlem1955, Wang1968}
had been reconsidered, corrected with the Stokes drift, and
compared with the experimental observations, which demonstrated a
good agreement between the theoretical and experimental results.
For high Reynolds numbers such that $Re_{s}=O(1)$
or $Re_{s}\gg 1$, the steady streaming induced by an
oscillating cylinder in an unbounded fluid had been studied in
\cite{Stuart1963,Stuart1966,Riley1965,Riley1967,Riley1975}.
For $Re_{s}=O(1)$, the steady flow outside the Stokes layer is
governed by the steady Navier-Stokes equations, and it had been
shown that for $Re_{s}\gg 1$ there is a double boundary layer near
the oscillating cylinder and that the steady streaming takes the
form of a jet-like flow along the axis of oscillations. In
\cite{HaddonRiley1979,Duck1979}, the steady flow between two
coaxial cylinders produced by small-amplitude transverse
oscillations of the inner cylinder had been studied in the case of
$Re_{s}=O(1)$ (and $Re_{s}\gg 1$). The results of \cite{Duck1979}
are a good agreement with the experimental results of \cite{Bertelsen1974}.
In \cite{HaddonRiley1979,Duck1979}, the case of small $Re_{s}$
corresponding to $\nu=O(1)$ had also been briefly discussed.

\vskip 3mm \noindent The aim of this paper is to investigate
steady streaming between two circular cylinders in the case of
$\nu=O(1)$. As was mentioned above, this problem had been treated before in
\cite{Duck1979}, where an asymptotic solution of the Navier-Stokes
equations had been constructed using the method of matched
asymptotic expansions. Formulae given in \cite{Duck1979} can be
used to write down a (composite) uniformly valid expansion
for the averaged stream function. For an oscillating cylinder in an
unbounded fluid, this
had been explicitly done  in \cite{Wang1968}. However, the asymptotic
solutions for both a single cylinder (presented in \cite{Wang1968})
and for two cylinders (presented in \cite{Duck1979}) are incomplete
because of the absence of the $O(\alpha^{3/2})$ term associated
with the averaged outer flow in the expansion of the averaged
stream function (this will be discussed in more detail in Section
5). In the present paper, we construct a uniformly valid expansion
of the averaged stream function in powers of $\alpha^{1/2}$ up to
terms of order $O(\alpha^{3/2})$.

\vskip 3mm \noindent
An interesting question which had not been discussed in \cite{Wang1968,Duck1979}
is the relation of the expansions obtained there to the original problem of a steady
steaming produced by an oscillating cylinder as it is seen by an observer fixed in space.
In \cite{Wang1968}, the problem of a steady streaming produced by a cylinder which is placed
in an oscillating flow had been considered. However, the steady component of the Eulerian velocity
depends on the reference frame that is used to describe the flow.
To clarify this point is one of the aims of this study.

\vskip 3mm \noindent
A similar problem arises in relation to the results of
\cite{Duck1979}, where
a conformal mapping that maps the gap between two eccentric cylinders
onto the  annulus between two cylinders with a common axis was employed.
The subsequent analysis had been done using
the transformed coordinates that depend on time, and the latter leads to a question whether
a transformation back to the physical coordinates will change the steady component
of the flow. The results of the present study give an answer to this question.

\vskip 3mm \noindent
Another question which is related to the previous two is the effect of the Stokes drift,
which is understood here as a difference between the averaged Eulerian velocity field and
the averaged Lagrangian velocity (the velocity of fluid particles). The importance of
Stokes drift is evident: (i) it is
the Lagrangian velocity (rather
than the Eulerian velocity) that is observed in experiments;
(ii) it is the Lagrangian velocity that is invariant under the change of the frame of reference
from the one fixed in the oscillating cylinder to the one fixed in space (and vice versa).
The Stokes drift in various oscillating flows had been studied before by many authors
(see, e.g., \cite{Longuet,Dore,Skavlem1955,Bertelsen1973}).
However, there is still no clarity about when one can expect a nonzero effect of the Stokes drift
in a flow produced by an oscillating cylinder. One of the aims of the present paper
is to resolve this ambiguity.

\vskip 3mm \noindent
To construct the asymptotic expansion of the
solution of the Navier-Stokes equations for small $\alpha$,
we employ the Vishik-Lyusternik
method\footnote {Nayfeh refers to this method as the method of
composite expansions \cite{Nayfeh}.} (see, e.g., \cite{Trenogin,
Nayfeh}) rather than the method of matched asymptotic expansions
which is routinely used in fluid mechanics. In most cases that we
know, at least first two terms in the uniformly valid
asymptotic expansions produced by these two methods are the same.
This does not mean that, given an expansion obtained by one of
the two methods, it is easy to derive the same expansion using the other one.
For example, although the asymptotic expansion constructed here using
the Vishik-Lyusternik method can be obtained by
the method of matched asymptotic expansions, this is not a straightforward
procedure (it requires a certain transformation of the velocity field
before the asymptotic procedure is started).
In comparison with the method of matched asymptotic expansions, the
Vishik-Lyusternik method involves more algebra in computing higher
order terms, but it has two essential advantages: (i) it does not
require the procedure of matching the inner and outer expansions
and (ii) the boundary layer part of the expansion
satisfies the condition of decay at infinity (in boundary layer variable)
in all orders of the expansion, which is not the case in
the method of matched asymptotic expansions where the boundary layer part
usually does not decay and may even grow at infinity.
The Vishik-Lyusternik method had been used to study
viscous boundary layers at a fixed impermeable boundary by Chudov
\cite{Chud}. Recently, it has been applied to viscous boundary
layers in high Reynolds number flows through a fixed domain with
an inlet and an outlet \cite{Ilin2008} and to viscous flows in a
half-plane produced by tangential vibrations on its boundary
\cite{VV2008}.

\vskip 3mm \noindent
In the present paper we compute first two nonzero terms  in the asymptotic expansion
of the steady velocity produced by an oscillating inner cylinder
and the corresponding expansion for the stream function.
The steady Eulerian velocity field is then corrected with the Stokes drift.
The results for an oscillating cylinder in an unbounded fluid
are obtained by passing to the limit $R\to\infty$ (where $R$ is the ratio of the radius of
the outer cylinder to the radius of the inner cylinder).

\vskip 3mm \noindent
The outline of the paper is as follows. In Section 2, we formulate the mathematical
problem. In Section 3, we describe
the method of constructing the asymptotic expansion and derive
the equations and boundary conditions that are to be solved.
These equations are solved and then corrected with the Stokes drift in Section 4.
Section 5 contains the discussion of the results.


\setcounter{equation}{0}
\renewcommand{\theequation}{2.\arabic{equation}}

\section{Formulation of the problem}

\vskip 3mm \noindent
We consider a two-dimensional flow of a viscous incompressible fluid between two circular cylinders
with radii $a$ and $b$ ($b > a$) produced by small translational vibrations of an inner cylinder about
the axis of the outer cylinder which is fixed in space.
Let $\bx^{*}=(x^{*},y^{*})$ be Cartesian coordinates on the plane and
let $\bx^{*}_{0}=(x^{*}_{0}(t^{*}), 0)$
be the position of the centre of the inner cylinder at time $t^{*}$.
We assume that
$x^{*}_{0}(t^{*})$ is
oscillating in $t^{*}$ with angular frequency $\omega$ and period
$T=2\pi/\omega$.
Using $1/\omega$, $a$, $V_{0}^{*}$ and $\rho a
\omega V_{0}^{*}$ as the characteristic scales for time, length,
velocity and pressure ($\rho$ is the fluid density), we introduce the dimensionless
variables $\tau=\omega \, t^{*}$, $\bx=\frac{1}{a}\bx^{*}$,
$\bx_{0}=\frac{1}{a}\bx^{*}_{0}$, $\bv=\frac{1}{V_{0}^{*}}\bv^{*}$ and
$p=\frac{p^{*}}{\rho a \omega V_{0}^{*}}$.
The motion of the fluid is
governed by the two-dimensional Navier-Stokes equations
\begin{equation}
\bv_{\tau}+\eps^2\left[(\bv\cdot\nabla)\bv-\nu
\nabla^{2}\bv\right]=-\nabla p, \quad \nabla\cdot\bv=0.
\label{2.4}
\end{equation}
Here $\eps=\sqrt{\alpha} \, $, $\alpha$ and $\nu$ are the inverse Strouhal
number and the inverse Reynolds number (the dimensionless
viscosity), defined by (\ref{1.1}). The velocity of the fluid
satisfies the standard no-slip condition on the surfaces of the
cylinders
\begin{equation}
\bv(\bx,\tau)\bigm\vert_{\rm outer \ cylinder}=0, \quad
\bv(\bx,\tau)\bigm\vert_{\rm inner \ cylinder}=\frac{1}{\eps^2}\dot{\bx}_{0}(\tau)=\frac{1}{\eps^2}\dot{x}_{0}(\tau)
\, \be_{x}. \label{2.2}
\end{equation}
Here dots denote differentiation with respect to $\tau$ and $x_{0}(\tau)$ is a given function which prescribes
the motion of the inner cylinder.
In what follows we are interested in
the asymptotic behaviour of periodic solution of Eqs. (\ref{2.4}), (\ref{2.2}) in the high-frequency limit $\eps\to 0$.
We assume that the amplitude of the oscillations of the cylinder is $O(\eps^2)$, i.e.
$x_{0}(\tau)=\eps^2 f(\tau)$
for some $2\pi$-periodic function $f$ and $f=O(1)$. In what follows, we will consider $f(\tau)$ given by
\begin{equation}
f(\tau)=Re\left(Ce^{i\tau}\right) \label{2.3}
\end{equation}
where $C$ is a complex constant having unit modulus ($\vert C\vert=1$).

\vskip 3mm
\noindent
Boundary conditions (\ref{2.2}) take the form
\begin{equation}
\bv(x,\tau)\bigm\vert_{\rm outer \ cylinder}=0, \quad
\bv(x,\tau)\bigm\vert_{\rm inner \ cylinder}=f^{\prime}(\tau)\, \be_{x}. \label{2.5}
\end{equation}
The time-dependent boundary of the inner cylinder can be described in the parametric form by the equations
\begin{equation}
x=\cos\tilde{\theta}+\epsilon^2 f(\tau), \quad y=\sin\tilde{\theta}  \label{2.6}
\end{equation}
where $\tilde{\theta}\in[0,2\pi)$ is the parameter on the cylinder boundary.
Now the boundary condition on the inner cylinder can be written as
\begin{equation}
\bv\!\biggm\vert_{
\stackrel{\scriptscriptstyle x=\cos\tilde{\theta}+\epsilon^2\! f(\tau)}
{\scriptscriptstyle y=\sin\tilde{\theta}}}
=f'(\tau) \, \be_{x}. \label{2.7}
\end{equation}
Using the assumption that $\epsilon$ is small, we expand $u$ and $v$ in Taylor's series at point
$(x,y)=(\cos\tilde{\theta}, \sin\tilde{\theta})$. This yields
\begin{equation}
\bv\!\biggm\vert_{
\stackrel{\scriptscriptstyle x=\cos\tilde{\theta}}
{\scriptscriptstyle y=\sin\tilde{\theta}}}+ \,
\epsilon^2 f(\tau) \, \pr_{x}\bv\!\biggm\vert_{
\stackrel{\scriptscriptstyle x=\cos\tilde{\theta}}
{\scriptscriptstyle y=\sin\tilde{\theta}}} \, +
\frac{\epsilon^4 f^2(\tau)}{2} \, \pr_{x}^{2}\bv\!\biggm\vert_{
\stackrel{\scriptscriptstyle x=\cos\tilde{\theta}}
{\scriptscriptstyle y=\sin\tilde{\theta}}}+\dots =f'(\tau) \, \be_{x}\label{2.8}
\end{equation}
Note that each term on the left side of Eq. (\ref{2.8}) is evaluated at the averaged position of the inner
cylinder (where the axes of both cylinders coincide).

\vskip 3mm
\noindent
In polar coordinates $(r,\theta)$ with origin at the axis of the outer cylinder,
Eqs. (\ref{2.4}) take the form
\begin{eqnarray}
&&u_{\tau}=-p_{r}+\eps^{2}\left[-uu_{r}-\frac{v}{r}u_{\theta}+\frac{v^{2}}{r}+
\nu \left(\nabla^{2}u-\frac{u}{r^{2}}-\frac{2}{r^{2}}v_{\theta} \right)\right], \quad \label{2.9} \\
&&v_{\tau}=-\frac{1}{r}p_{\theta}+\eps^{2}\left[-uv_{r}-\frac{v}{r}v_{\theta}-\frac{u v}{r}+
\nu \left(\nabla^{2}v-\frac{v}{r^{2}}+\frac{2}{r^{2}}u_{\theta} \right)\right], \quad \label{2.10} \\
&&u_{r}+\frac{u}{r}+\frac{1}{r}v_{\theta}=0, \label{2.11}
\end{eqnarray}
where $u$ and $v$ are the radial and azimuthal components of the velocity, subscripts `$\tau$', `$r$' and
`$\theta$' denote partial derivatives, and where
$\nabla^{2}=\pr_{r}^{2} + (1/r)\pr_{r}+(1/r^{2})\pr_{\theta}^{2}$.
The boundary conditions at the outer cylinder are
\begin{equation}
u\Bigm\vert_{r=R}=0, \quad v\Bigm\vert_{r=R}=0  \label{2.12}
\end{equation}
where $R=b/a$. The boundary condition (\ref{2.8}) at the inner cylinder takes the form
\begin{eqnarray}
u\!\Bigm\vert_{r=1} &+&
\epsilon^2 f \left[Lu
+\frac{\sin\theta}{r} v\right]\!\biggm\vert_{r=1} \nonumber \\
&+&\frac{\epsilon^4 f^2}{2} \left[L^2u - \frac{\sin^2\theta}{r}u +\frac{2\sin\theta}{r}L v -
\frac{2\sin\theta\cos\theta}{r^2}v\right]
\!\biggm\vert_{r=1}
+\dots \, =f'(\tau)\cos\theta, \quad\quad \label{2.13} \\
v\!\Bigm\vert_{r=1} &+&
\epsilon^2 f \left[Lv
-\frac{\sin\theta}{r}u \right]\!\biggm\vert_{r=1} \nonumber \\
&+&\frac{\epsilon^4 f^2}{2} \left[L^2v - \frac{\sin^2\theta}{r}v -\frac{2\sin\theta}{r}L u +
\frac{2\sin\theta\cos\theta}{r^2}u\right]
\!\biggm\vert_{r=1}+\dots \, =-f'(\tau)\sin\theta . \label{2.14}
\end{eqnarray}
Here $L=\cos\theta \, \pr_{r}-(\sin\theta /r) \, \pr_{\theta}$.


\setcounter{equation}{0}
\renewcommand{\theequation}{3.\arabic{equation}}

\section{Asymptotic expansion}

\vskip 3mm
\noindent
We seek a solution of (\ref{2.9})--(\ref{2.14}) in the form
\begin{eqnarray}
&&u=u^{i}(r,\theta,\tau,\eps)+\eps \, u^{a}(\xi,\theta,\tau,\eps)+\eps \, u^{b}(\eta,\theta,\tau,\eps), \label{3.1} \\
&&v=v^{i}(r,\theta,\tau,\eps)+v^{a}(\xi,\theta,\tau,\eps)+v^{b}(\eta,\theta,\tau,\eps), \label{3.2} \\
&&p=p^{i}(r,\theta,\tau,\eps)+p^{a}(\xi,\theta,\tau,\eps)+ p^{b}(\eta,\theta,\tau,\eps). \label{3.3}
\end{eqnarray}
Here $\xi=(r-1)/\epsilon$ and $\eta=(R-r)/\epsilon$ are the boundary layer variables.
Functions $u^{i}$, $v^{i}$, $p^{i}$ represent
a regular expansion of the solution in power series in $\eps$ (an outer solution), and
$u^{a}$, $v^{a}$, $p^{a}$ and $u^{b}$, $v^{b}$, $p^{b}$ correspond to boundary layer corrections
to this regular expansion. Superscripts `a' and `b' correspond to the boundary layers at the inner and
outer cylinders respectively.
We assume that the boundary layer part of the expansion rapidly decays outside thin boundary layers,
i.e. $u^{a}, v^{a}, p^{a} \to 0$ as $\xi\to\infty$ and $u^{b}, v^{b}, p^{b} \to 0$ as $\eta\to\infty$.

\subsection{Regular part of the expansion}

\noindent
To derive the equations governing the regular part of the expansion,
it is convenient to work with the Cartesian form (\ref{2.4}) of the Euler equations.
Later we can rewrite the results in polar coordinates.
Let
\begin{equation}
\bv^{i}=\bv^{i}_{0}+ \eps \, \bv^{i}_{1}+\eps^2 \bv^{i}_{2}+ \dots , \quad
p^{i}=p^{i}_{0}+ \eps \, p^{i}_{1}+\eps^2 p^{i}_{2}+ \dots , \label{3.4}
\end{equation}
where $\bv^{i}=v^{i}_{1}\mathbf{e}_{x}+v^{i}_{2}\mathbf{e}_{y}$ and
$\bv^{i}_{k}=v^{i}_{1k}\mathbf{e}_{x}+v^{i}_{2k}\mathbf{e}_{y}$ ($k=0,1,2,\dots$).
On substituting (\ref{3.4}) in (\ref{2.4}) and collecting terms of equal powers of $\eps$, we find that
the successive approximations $\bv^{i}_{k}$, $p^{i}_{k}$ ($k=0,1,2,\dots$)
satisfy the equations:
\begin{equation}
\pr_{\tau}\bv^{i}_{k}=-\nabla p^{i}_{k}, \quad \nabla\cdot\bv^{i}_{k}=0 \label{3.5}
\end{equation}
for $k=0,1$ and
\begin{equation}
\pr_{\tau}\bv^{i}_{k}=-\nabla p^{i}_{k}+
\left(-\sum_{l=0}^{k-2}(\bv^{i}_{l}\cdot\nabla)\bv^{i}_{k-2-l}
+\nu\nabla^{2} \bv^{i}_{k-2}\right), \quad
\nabla\cdot\bv^{i}_{k}=0. \label{3.6}
\end{equation}
for $k\geq 2$.
In what follows, we will use the following notation: for any $2\pi$-periodic function $F(\tau)$,
\[
F(\tau)=\bar{F}+\tilde{F}(\tau), \quad \bar{F}=\frac{1}{2\pi}\int\limits_{0}^{2\pi}F(\tau)d\tau,
\]
i.e. $\bar{F}$ is the mean value of $F(\tau)$ and, by definition,
$\tilde{F}(\tau)=F(\tau)-\bar{F}$ is the oscillating part of $F(\tau)$.

\vskip 3mm
\noindent
{\em Leading-order terms.} Consider first Eqs. (\ref{3.5}) for $k=0$. We seek a
solution $\bv^{i}_{0}$ which is periodic in $\tau$. Averaging the
equation for $\bv^{i}_{0}$ yields: $\nabla\bar{p}^{i}_{0}=0$, which
is the necessary condition for existence of periodic (in $\tau$)
solutions for $\bv^{i}_{0}$. Without loss of generality, we put $\bar{p}^{i}_{0}=0$,
i.e. in the leading order the pressure in the outer solution is purely oscillatory with zero mean value.
The general solution of Eqs. (\ref{3.5}) can be written as
$\bv^{i}_{0}=\bar{\bv}^{i}_{0}+\tilde{\bv}_{0}$ where $\tilde{\bv}^{i}_{0}=\nabla\phi_{0}$ and
$\phi_{0}$ has zero mean value and is the solution of the boundary value problem
\begin{equation}
\nabla^{2}\phi_{0}=0, \quad
\phi_{0r}\!\!\bigm\vert_{r=1}=f'(\tau)\cos\theta, \quad
\phi_{0r}\!\!\bigm\vert_{r=R}= 0. \label{3.8}
\end{equation}
The boundary conditions for $\phi_{0r}$ at $r=1$ and $r=R$ will be justified later.

\vskip 3mm
\noindent
On averaging the equations for $\bv^{i}_{2}$ and the second of equations (\ref{3.5}), we obtain
\begin{equation}
(\bar{\bv}^{i}_{0}\cdot\nabla)\bar{\bv}^{i}_{0}
+\overline{(\tilde{\bv}^{i}_{0}\cdot\nabla)\tilde{\bv}^{i}_{0}}
=-\nabla \bar{p}^{i}_{2}
+\nu\nabla^{2} \bar{\bv}^{i}_{0}, \quad
\nabla\cdot\bar{\bv}^{i}_{0}=0. \label{3.9}
\end{equation}
Since $\tilde{\bv}^{i}_{0}$ is irrotational, we can rewrite (\ref{3.9}) as
\begin{equation}
(\bar{\bv}^{i}_{0}\cdot\nabla)\bar{\bv}^{i}_{0}
=-\nabla \Pi_{0}
+\nu\nabla^{2} \bar{\bv}^{i}_{0}, \quad
\nabla\cdot\bar{\bv}^{i}_{0}=0, \label{3.10}
\end{equation}
where $\Pi_{0}=\bar{p}^{i}_{2}+\overline{\vert\nabla\phi_{0}\vert^2/2}$.
Equations (\ref{3.10}) represent the time-independent Navier-Stokes equations that describe
steady flows of a viscous incompressible fluid.
Boundary conditions for $\bar{\bv}^{i}_{0}$ will be specified later.

\vskip 3mm
\noindent
{\em First-order terms.} The solution of Eqs. (\ref{3.5}) for $k=1$ has the form
$\bv^{i}_{1}=\bar{\bv}^{i}_{1}+\tilde{\bv}_{1}$ where $\tilde{\bv}^{i}_{1}=\nabla\phi_{1}$
and $\phi_{1}$ has zero mean value and is the solution of the boundary value problem
\begin{equation}
\nabla^{2}\phi_{1}=0, \quad
\phi_{1r}\!\!\bigm\vert_{r=1}=a_{1}(\theta,\tau), \quad
\phi_{1r}\!\!\bigm\vert_{r=R}=b_{1}(\theta,\tau). \label{3.12}
\end{equation}
Functions $a_{1}(\theta,\tau)$ and $b_{1}(\theta,\tau)$ will be defined later.
Manipulations similar to those employed in derivation of Eqs. (\ref{3.10}) lead to the following equations
for $\bar{\bv}^{i}_{1}$:
\begin{equation}
(\bar{\bv}^{i}_{0}\cdot\nabla)\bar{\bv}^{i}_{1}
+(\bar{\bv}^{i}_{1}\cdot\nabla)\bar{\bv}^{i}_{0}
=-\nabla \Pi_{1}
+\nu\nabla^{2} \bar{\bv}^{i}_{1}, \quad
\nabla\cdot\bar{\bv}^{i}_{1}=0, \label{3.13}
\end{equation}
where $\Pi_{1}=\bar{p}^{i}_{3}+\overline{\left(\nabla\phi_{0}\cdot\nabla\phi_{1}\right)}$.
Boundary conditions for $\bar{\bv}^{i}_{1}$ will be specified later.

\vskip 3mm
\noindent
{\em Second-order terms.} It will be proved later that
\begin{equation}
\bar{\bv}^{i}_{0}\equiv 0 \quad {\rm and} \quad \bar{\bv}^{i}_{1}\equiv 0. \label{3.14}
\end{equation}
Using these and the fact that both $\tilde{\bv}^{i}_{0}$ and $\tilde{\bv}^{i}_{1}$ are irrotational,
it can be shown that
\[
\pr_{\tau}\tilde{\bv}^{i}_{2}=-\nabla Q_{2}, \quad
\nabla\cdot\tilde{\bv}^{i}_{2}=0,
\]
where $Q_{2}=\tilde{p}^{i}_{2}+\frac{(\nabla\phi_{0})^{2}}{2}-\frac{\overline{(\nabla\phi_{0})^{2}}}{2}$. It follows that
$\tilde{\bv}^{i}_{2}$ is irrotational, i.e. $\tilde{\bv}^{i}_{2}=\nabla\phi_{2}$, and
$\phi_{2}$ is the solution of the boundary value problem
\begin{equation}
\nabla^{2}\phi_{2}=0, \quad
\phi_{2r}\!\!\bigm\vert_{r=1}=a_{2}(\theta,\tau), \quad
\phi_{2r}\!\!\bigm\vert_{r=R}=b_{2}(\theta,\tau), \label{3.15}
\end{equation}
where functions $a_{2}(\theta,\tau)$ and $b_{2}(\theta,\tau)$ will be defined later.

\vskip 3mm
\noindent
The equations for $\bar{\bv}^{i}_{2}$ can be written in the form
\begin{equation}
0=-\nabla \Pi_{2}
+\nu\nabla^{2} \bar{\bv}^{i}_{2}, \quad
\nabla\cdot\bar{\bv}^{i}_{2}=0, \label{3.16}
\end{equation}
where $\Pi_{2}=\bar{p}^{i}_{4}+\overline{\nabla\phi_{0}\cdot\nabla\phi_{2}}+
\overline{\frac{(\nabla\phi_{1})^2}{2}}$ and where we have used the assumptions (\ref{3.14}).
Thus, the second order averaged velocity is described by
the Stokes equations. Again, boundary conditions for $\bar{\bv}^{i}_{2}$ will be specified later.

\vskip 3mm
\noindent
{\em Third-order terms.} Separating the oscillating part of the equation for $\bv^{i}_{3}$ and employing
(\ref{3.14}) and the fact that $\tilde{\bv}^{i}_{0}$ and
$\tilde{\bv}^{i}_{1}$ are irrotational, we find that
\[
\pr_{\tau}\tilde{\bv}^{i}_{3}=-\nabla Q_{3},
\]
where $Q_{3}=\tilde{p}^{i}_{3}+\nabla\phi_{0}\cdot\nabla\phi_{1}-\nabla\phi_{0}\cdot\nabla\phi_{1}$.
This equation and the continuity equation $\nabla\cdot\tilde{\bv}^{i}_{3}=0$ imply that
$\tilde{\bv}^{i}_{3}=\nabla\phi_{3}$
where $\phi_{3}$ is the solution of the boundary value problem
\begin{equation}
\nabla^{2}\phi_{3}=0, \quad
\phi_{3r}\!\!\bigm\vert_{r=1}=a_{3}(\theta,\tau), \quad
\phi_{3r}\!\!\bigm\vert_{r=R}=b_{3}(\theta,\tau). \label{3.18}
\end{equation}
Function $a_{3}(\theta,\tau)$ and $b_{3}(\theta,\tau)$ will be defined later.

\vskip 3mm
\noindent
The equations for $\bar{\bv}^{i}_{3}$ can be derived in the same manner as the equations
for $\bar{\bv}^{i}_{0}$, $\bar{\bv}^{i}_{1}$ and $\bar{\bv}^{i}_{2}$. They are given by
\begin{equation}
0=-\nabla \Pi_{3}
+\nu\nabla^{2} \bar{\bv}^{i}_{3}, \quad
\nabla\cdot\bar{\bv}^{i}_{3}=0, \label{3.19}
\end{equation}
where $\Pi_{3}=\bar{p}^{i}_{5}+\overline{\nabla\phi_{0}\cdot\nabla\phi_{3}}+
\overline{\nabla\phi_{1}\cdot\nabla\phi_{2}}$.

\vskip 3mm
\noindent
Thus, we have found that both the second and third order averaged
velocities ($\bar{\bv}^{i}_{2}$ and $\bar{\bv}^{i}_{3}$)
in the outer flow are solutions of the Stokes problem with boundary conditions which will be determined later.


\vskip 5mm
\noindent
\subsection{Boundary layers}

\vskip 3mm
\noindent
{\it Boundary layer at the inner cylinder}. We assume that
\begin{equation}
u^{a}=u^{a}_{0}+ \eps \, u^{a}_{1}+ \dots , \ \
v^{a}=v^{a}_{0}+ \eps \, v^{a}_{1}+  \dots , \ \
p^{a}=p^{a}_{0}+ \eps \, p^{a}_{1}+  \dots . \label{3.20}
\end{equation}
Now we use our assumption that $u^{b}$, $v^{b}$ and $p^{b}$ are nonzero only within a thin boundary
layer near the outer cylinder and drop them from Eqs. (\ref{3.1})--(\ref{3.3}). Then we
substitute the resulting equations,
as well as (\ref{3.4}) and (\ref{3.20}), into Eqs.
(\ref{2.9})--(\ref{2.11})  and take into account that $u^{i}_{k}$, $v^{i}_{k}$, $p^{i}_{k}$
($k=0,1,\dots$)
satisfy the equations (\ref{3.5}), (\ref{3.6}). After that,
we make the change of variables
$r=1+\eps \, \xi$, expand every function of $\eps \, \xi$ in Taylor's series at $\eps=0$
and collect terms of equal powers in $\eps$. This yields the following equations:
\begin{eqnarray}
&&v^{a}_{k\tau}-\nu v^{a}_{k\xi\xi}+p^{a}_{k\theta}=F^{a}_{k}, \label{3.24} \\
&&p^{a}_{k\xi}=G^{a}_{k}, \label{3.25} \\
&&u^{a}_{k\xi}+v^{a}_{k\theta}=H^{a}_{k}, \label{3.26}
\end{eqnarray}
for $k=0,1,\dots$
Here $F^{a}_{0}=0$, $G^{a}_{0}=0$ and $H^{a}_{0}=0$; for $k > 0$,
functions $F^{a}_{k}$, $G^{a}_{k}$ and $H^{a}_{k}$ are defined in term of $\bv^{i}_{0}, \dots, \bv^{i}_{k-1}$,
$v^{a}_{0}, \dots, v^{a}_{k-1}$ and $u^{a}_{0}, \dots, u^{a}_{k-1}$. Explicit expressions for these functions
are given in Appendix A (for $k=1,2,3$).

\vskip 3mm
\noindent
{\it Boundary layer at the outer cylinder}. Let
$u^{b}=u^{b}_{0}+ \eps \, u^{b}_{1}+ \dots$,
$v^{b}=v^{b}_{0}+ \eps \, v^{b}_{1}+ \dots$ and
$p^{b}=p^{b}_{0}+ \eps \, p^{b}_{1}+ \dots$
The same procedure as before
produces the following sequence of equations:
\begin{eqnarray}
&&v^{b}_{k\tau}-\nu v^{b}_{k\eta\eta}+\frac{1}{R}p^{b}_{k\theta}=F^{b}_{k}, \label{3.31} \\
&&p^{b}_{k\eta}=G^{b}_{k}, \label{3.32} \\
&&-u^{b}_{k\eta}+\frac{1}{R}v^{b}_{k\theta}=H^{b}_{k}, \label{3.33}
\end{eqnarray}
for $k=0,1,\dots$
Here $F^{b}_{0}=0$, $G^{b}_{0}=0$ and $H^{b}_{0}=0$; explicit expressions for
$F^{b}_{k}$, $G^{b}_{k}$ and $H^{b}_{k}$
are given in Appendix A (for $k=1,2,3$).

\vskip 3mm
\noindent
We require that in all orders the boundary layer corrections to the outer solution rapidly decay
outside boundary layers, i.e. (for each $k=0,1,\dots$)
\begin{equation}
u^{a}_{k}, \ v^{a}_{k}, \ p^{a}_{k}\to 0  \quad {\rm as} \quad \xi\to\infty
\quad {\rm and} \quad u^{b}_{k}, \ v^{b}_{k}, \ p^{b}_{k}\to 0 \quad {\rm as} \quad \eta\to\infty . \label{3.34}
\end{equation}


\vskip 5mm
\noindent
\subsection{Boundary conditions}

\vskip 3mm
\noindent
Now we take our expansions of the velocity in the outer flow and in the boundary layers, substitute them into (\ref{2.12})--(\ref{2.14})
and collect terms of equal powers in $\eps$. This produces the following boundary conditions:
\begin{eqnarray}
u^{i}_{k}\!\!\bigm\vert_{r=1}+ u^{a}_{k-1}\!\!\bigm\vert_{\xi=0} &=& Q_{k} , \label{3.40} \\
v^{i}_{k}\!\!\bigm\vert_{r=1}+v^{a}_{k}\!\!\bigm\vert_{\xi=0} &=&S_{k}, \label{3.41} \\
u^{i}_{k}\!\!\bigm\vert_{r=R}+ u^{b}_{k-1}\!\!\bigm\vert_{\eta=0}&=& 0, \label{3.42} \\
v^{i}_{k}\!\!\bigm\vert_{r=R}+v^{b}_{k}\!\!\bigm\vert_{\eta=0} &=& 0. \label{3.43}
\end{eqnarray}
for $k=0,1,\dots$
Here
\[
Q_{0}=f'(\tau) \cos\theta, \quad
S_{0}=- f'(\tau) \sin\theta ,
\]
functions
$Q_{k}$ and $S_{k}$ for $k > 0$ depend on $\bv^{i}_{0}, \dots, \bv^{i}_{k-1}$,
$v^{b}_{0}, \dots, v^{b}_{k-1}$ and $u^{b}_{0}, \dots, u^{b}_{k-1}$ and are given in Appendix A
(for $k=1,2,3$).
Note that boundary conditions for $\phi_{0r}$ at $r=1$ and $r=R$ in the boundary value problem
(\ref{3.8}) follow directly from (\ref{3.40}) and (\ref{3.42}) with $k=0$.


\setcounter{equation}{0}
\renewcommand{\theequation}{4.\arabic{equation}}

\section{Analysis of the asymptotic equations}


\subsection{Leading order equations}

\vskip 3mm
\noindent
{\it Outer flow.} The solution of (\ref{3.8}) that describes the (leading order) oscillating outer
flow is
\begin{equation}
\phi_{0}=- \frac{f^{\prime}(\tau)}{R^2-1}\left(r+\frac{R^2}{r}\right)\cos\theta. \label{4.1}
\end{equation}
{\it Inner cylinder.} Consider now Eqs. (\ref{3.24})--(\ref{3.26}) for $k=0$.
The condition of decay at infinity (in variable $\xi$) for $p^{a}_{0}$ and Eq. (\ref{3.25})
have a consequence that $p^{a}_{0}\equiv 0$. Equation (\ref{3.24}) simplifies to the standard heat equation
\begin{equation}
v^{a}_{0\tau}=\nu v^{a}_{0\xi\xi}. \label{4.2}
\end{equation}
Boundary condition for $v^{a}_{0}$ at $\xi=0$ follows from (\ref{3.41}) (with $k=0$):
\begin{equation}
v^{a}_{0}\!\bigm\vert_{\xi=0} =
-v^{i}_{0}\!\bigm\vert_{r=1} - f^{\prime}(\tau) \sin\theta =
-\frac{2R^2}{R^2-1}Re\left(iCe^{i\tau}\right) \sin\theta. \label{4.3}
\end{equation}
The solution of (\ref{4.2}) subject to
the boundary conditions (\ref{4.3}) and (\ref{3.34}) is given by
\begin{equation}
v^{a}_{0}=
-\frac{2R^2}{R^2-1}Re\left(iCe^{-\gamma\xi+i\tau}\right) \sin\theta, \quad
\gamma=\frac{1+i}{\sqrt{2\nu}}. \label{4.4}
\end{equation}
It follows from (\ref{4.4}) that $\bar{v}^{a}_{0}=0$.
Thus, {\em in the leading order
the boundary layer at the inner cylinder is a purely oscillatory Stokes layer}. This fact implies that
the boundary condition for $\bar{v}^{i}_{0}$ at $r=1$ (that
is obtained by averaging the condition (\ref{3.41})) is $\bar{v}^{i}_{0}\!\!\bigm\vert_{r=1}=0$.
Similarly, averaging the condition (\ref{3.40}) yields $\bar{u}^{i}_{0}\!\!\bigm\vert_{r=1}=0$.
Thus, we have
\begin{equation}
\bar{\bv}^{i}_{0}\!\bigm\vert_{r=1}=0. \label{4.5}
\end{equation}

\vskip 3mm
\noindent
The normal velocity $u^{a}_{0}$ is determined from
Eq. (\ref{3.26}):
\begin{equation}
u^{a}_{0}(\xi,\theta,\tau)=\int\limits_{\xi}^{\infty}v^{a}_{0\theta}(\xi',\theta,\tau) \, d\xi'
=-\frac{2R^2}{R^2-1}Re\left(\frac{i}{\gamma}Ce^{-\gamma\xi+i\tau}\right) \cos\theta . \label{4.6}
\end{equation}
Here the constant of integration is chosen so as to guarantee that $u^{a}_{0}(\xi,\theta,\tau)$ decays as
$\xi\to\infty$. $u^{a}_{0}\!\!\bigm\vert_{\xi=0}$ gives us the boundary condition for
the next approximation of the outer solution. Indeed, according to (\ref{3.40}) for $k=1$,
we must have
\begin{equation}
u^{i}_{1}(r,\theta,\tau)\!\bigm\vert_{r=0}=- u^{a}_{0}(\xi,\theta,\tau)\!\bigm\vert_{\xi=0}=
\frac{2R^2}{R^2-1}Re\left(\frac{i}{\gamma}Ce^{i\tau}\right) \cos\theta, \label{4.7}
\end{equation}
This equation defines function $a_{1}(\theta,\tau)$ in (\ref{3.12}).


\vskip 3mm
\noindent
{\it Outer cylinder.}
Consider now Eqs. (\ref{3.31})--(\ref{3.33}) for $k=0$. An analysis similar to what we did for the
boundary layer at the inner cylinder  results in the formula
\begin{equation}
v^{b}_{0}=
-\frac{2}{R^2-1}Re\left(iCe^{-\gamma\eta+i\tau}\right) \sin\theta . \label{4.11}
\end{equation}
As before, the radial velocity $u^{b}_{0}$ is determined from the incompressibility condition
(\ref{3.33}):
\begin{equation}
u^{b}_{0}=-\frac{1}{R}\int\limits_{\eta}^{\infty}v^{b}_{0\theta}(\eta',\theta,\tau) \, d\eta'
=\frac{2}{R(R^2-1)}Re\left(\frac{i}{\gamma}Ce^{-\gamma\xi+i\tau}\right) \cos\theta . \label{4.12}
\end{equation}
Again, the constant of integration is chosen so as to guarantee the decay of $u^{b}_{0}$ as
$\eta\to\infty$. $u^{b}_{0}\!\!\bigm\vert_{\eta=0}$ gives us the boundary condition for
the next approximation of the outer solution:
\begin{equation}
u^{i}_{1}\!\bigm\vert_{r=1}=- u^{b}_{0}\!\bigm\vert_{\eta=0}=
-\frac{2}{R(R^2-1)}Re\left(\frac{i}{\gamma}Ce^{i\tau}\right) \cos\theta. \label{4.13}
\end{equation}
This equation defines function $b_{1}(\theta,\tau)$ in (\ref{3.12}).

\vskip 3mm
\noindent
It follows from (\ref{4.11}) that $\bar{v}^{b}_{0}=0$, i.e.
{\it in the leading order
the boundary layer at the outer cylinder is purely oscillatory}. This, in turn, implies that
the boundary condition for $\bar{v}^{i}_{0}$ at $r=R$ (obtained
by averaging the condition (\ref{3.43})) is $\bar{v}^{i}_{0}\!\!\bigm\vert_{r=R}=0$.
Similarly, Eq. (\ref{3.42}) yields $\bar{u}^{i}_{0}\!\!\bigm\vert_{r=R}=0$.
Hence,
\begin{equation}
\bar{\bv}^{i}_{0}\!\bigm\vert_{r=R}=0. \label{4.15}
\end{equation}

\vskip 3mm
\noindent
{\it Averaged outer flow.}
Equations (\ref{3.10}) and boundary conditions (\ref{4.5}) and (\ref{4.15}) imply that
$\bar{\bv}^{i}_{0}\equiv 0$, i.e {\it there is no steady streaming
in the leading order of the expansion}. This justifies our earlier
assumption about $\bar{\bv}^{i}_{0}$.


\vskip 5mm
\noindent
\subsection{First order equations}

\vskip 3mm
\noindent
{\it Oscillatory outer flow.} Since now we know
functions $a_{1}(\theta,\tau)$ and $b_{1}(\theta,\tau)$ (defined by Eqs. (\ref{4.7}) and (\ref{4.13})),
we can solve problem (\ref{3.12}). The solution is given by
\begin{equation}
\phi_{1}=- \frac{2R}{(R^2-1)^2}\left((R+1)r+\frac{R^3+1}{r}\right)
Re\left(\frac{i C}{\gamma}e^{i\tau}\right)\cos\theta. \label{4.16}
\end{equation}
{\it Inner cylinder.}
Consider Eqs. (\ref{3.24})--(\ref{3.26}) for $k=1$.
The fact that $G^{a}_{1}=0$ (Eq. (\ref{A4}) in Appendix A) and the same arguments as before lead us
to conclusion that $p^{a}_{1}\equiv 0$. Hence, Eq. (\ref{3.24}) reduces to
\begin{equation}
v^{a}_{1\tau}-\nu v^{a}_{1\xi\xi}=-\cos\theta \, f^{\prime}(\tau)v^{a}_{0\xi} +\nu v^{a}_{0\xi}. \label{4.18}
\end{equation}
Averaging in $\tau$ and integrating in variable $\xi$ twice, we find that
\begin{equation}
\bar{v}^{a}_{1}=-\frac{1}{\nu} \, \cos\theta \,
\int\limits_{\xi}^{\infty}\overline{f^{\prime}(\tau)v^{a}_{0}} \, d\xi'. \label{4.19}
\end{equation}
Here the constants of integration are chosen so as to ensure that
$\bar{v}^{a}_{1}\to 0$ as $\xi\to\infty$.
It follows from the definition of averaging that
$\overline{f_{1}^{\prime}(\tau)f_{2}(\tau)}=-\overline{f_{1}(\tau)f_{2}^{\prime}(\tau)}$
for any $2\pi$-periodic functions $f_{1}$ and $f_{2}$. Employing this property in (\ref{4.19}), we obtain
\begin{equation}
\bar{v}^{a}_{1} =  \frac{1}{\nu} \, \cos\theta \,
\int\limits_{\xi}^{\infty}\overline{f(\tau)v^{a}_{0\tau}} \, d\xi' =  \cos\theta \,
\int\limits_{\xi}^{\infty}\overline{f(\tau)v^{a}_{0\xi\xi}} \, d\xi' = -  \cos\theta \,
\overline{f(\tau)v^{a}_{0\xi}} . \label{4.20}
\end{equation}
Here we used the fact that $v^{a}_{0}$ satisfies Eq. (\ref{4.2}).
Substitution of (\ref{2.3}) and (\ref{4.4}) in (\ref{4.20}) yields
\begin{equation}
\bar{v}^{a}_{1} =  \frac{1}{2\sqrt{2\nu}} \,  \frac{R^2}{R^2-1} \, e^{-\xi/\sqrt{2\nu}}
\left(\cos (\xi/\sqrt{2\nu})-\sin (\xi/\sqrt{2\nu})\right) \sin 2\theta . \label{4.21}
\end{equation}
The oscillatory part $\tilde{v}^{a}_{1}$ of ${v}^{a}_{1}$ satisfies the equation
\begin{equation}
\tilde{v}^{a}_{1\tau}-\nu \tilde{v}^{a}_{1\xi\xi}=-\cos\theta \,
\widetilde{f^{\prime}v^{a}_{0\xi}} +\nu v^{a}_{0\xi}, \label{4.22}
\end{equation}
the condition of decay at infinity and the boundary condition
\begin{equation}
\tilde{v}^{a}_{1}\!\!\bigm\vert_{\xi=0} = -
\frac{2R(R^3+R+2)}{(R^2-1)^2}Re\left(\frac{i C}{\gamma}e^{i\tau}\right)\sin\theta
- \cos\theta \, \widetilde{f v^{a}_{0\xi}}\!\!\Bigm\vert_{\xi=0}, \label{4.23}
\end{equation}
which follows from the oscillatory part of (\ref{3.41}) and from Eq. (\ref{4.16}).
Standard but tedious calculations result in
\begin{equation}
\tilde{v}^{a}_{1} =-\cos\theta \, \widetilde{f v^{a}_{0\xi}}+w, \quad
w = -\frac{2R^2}{R^2-1} \,
Re\left[i C\left(\frac{R^3+R+2}{R(R^2-1)} \, \frac{1}{\gamma} - \frac{\xi}{2}\right)e^{-\gamma\xi+i\tau}
\right] \sin\theta. \label{4.24}
\end{equation}
We do not give an explicit formula for $\widetilde{f v^{a}_{0\xi}}$ as we do not use it
in what follows\footnote{$\widetilde{f v^{a}_{0\xi}}$ as a function of $\tau$ is purely
oscillatory with double frequency of
oscillation and therefore, it produces zero contribution to all quantities
which will be  of interest to us.}. Both $\bar{u}^{a}_{1}$ and $\tilde{u}^{a}_{1}$ are computed
using Eq. (\ref{3.26}). We have
\begin{eqnarray}
&&\bar{u}^{a}_{1} =  \int\limits_{\xi}^{\infty}\bar{v}^{a}_{1\theta} \, d\xi =
-  \frac{R^2}{R^2-1} \, e^{-\xi/\sqrt{2\nu}}
\sin (\xi/\sqrt{2\nu}) \cos 2\theta, \label{4.26} \\
&&\tilde{u}^{a}_{1} = \pr_{\theta}\left(\cos\theta \, \widetilde{f v^{a}_{0}}\right)+h,  \quad
h = -\frac{2R^2}{R^2-1} \,
Re\left[\frac{i C}{\gamma}\left(\frac{R^3+3R+4}{R(R^2-1)} \, \frac{1}{2\gamma} -
\frac{3\xi}{2}\right)e^{-\gamma\xi+i\tau}
\right] \cos\theta. \quad\quad\quad\quad \label{4.27}
\end{eqnarray}

\noindent
{\it Outer cylinder.}
Similar analysis, applied to Eqs. (\ref{3.31})--(\ref{3.33}), yields
\begin{eqnarray}
&&v^{b}_{1} = -\frac{2\sin\theta}{R(R^2-1)} \,
Re\left[i C\left(\frac{2R^3+R^2+1}{R^2-1} \, \frac{1}{\gamma} + \frac{\eta}{2}\right)e^{-\gamma\eta+i\tau}
\right] , \label{4.29} \\
&&u^{b}_{1} =
\frac{2\cos\theta}{R^2(R^2-1)} \, Re\left[\frac{i C}{\gamma}\left(\frac{4R^3+3R^2+1}{R^2-1}\frac{1}{2\gamma}
+\frac{3\eta}{2}\right)e^{-\gamma\eta+i\tau}\right] . \quad\quad\quad \label{4.30}
\end{eqnarray}
Equations (\ref{4.29}) and (\ref{4.30}) imply that
$\bar{v}^{b}_{1}\equiv 0$ and $\bar{u}^{b}_{1}\equiv 0$,
i.e. in contrast with the inner cylinder, {\it there is no first-order steady
boundary layer at the outer cylinder}.

\vskip 3mm
\noindent
{\it Averaged outer flow.} On averaging boundary conditions (\ref{3.42}) and (\ref{3.43}) (for $k=1$)
and using the fact that $\bar{u}^{b}_{0}=0$ and $\bar{v}^{b}_{1}=0$, we find that
\begin{equation}
\bar{\bv}^{i}_{1}\!\!\bigm\vert_{r=R}=0. \label{4.31}
\end{equation}
Further, averaging boundary conditions (\ref{3.40}) and (\ref{3.41})
and using (\ref{4.6}) and (\ref{4.20}), we obtain
\[
\bar{u}^{i}_{1}\!\!\bigm\vert_{r=1} = - \bar{u}^{a}_{0}\!\!\bigm\vert_{\xi=0} = 0 , \quad
\bar{v}^{i}_{1}\!\!\bigm\vert_{r=1} = - \bar{v}^{a}_{1}\!\!\bigm\vert_{\xi=0}
- \cos\theta \, \overline{f(\tau) v^{a}_{0\xi}}\!\!\bigm\vert_{\xi=0}=0 .
\]
These, together with (\ref{4.31}), imply that Eqs. (\ref{3.13}) should be solved
with zero boundary conditions, which, in turn, leads  to a conclusion that
$\bar{\bv}^{i}_{1}\equiv 0$. This justifies our earlier assumption and means that {\it there is no steady outer flow
in the first order of the expansion}.


\vskip 5mm
\noindent
\subsection{Second order equations}

\vskip 3mm
\noindent
{\it Oscillatory outer flow.} To find the
oscillatory part of the second-order outer flow we need to solve problem
(\ref{3.15}) for the velocity potential $\phi_{2}$.
Functions
$a_{2}(\theta,\tau)$ and $b_{2}(\theta,\tau)$ which appear in (\ref{3.15})
are determined by boundary conditions (\ref{3.40}) and (\ref{3.42}) for $k=2$.
With the help of Eq. (\ref{3.40}) with $k=0$ and the continuity equations
for $\bv^{a}_{0}$ and $\bv^{i}_{0}$, boundary condition (\ref{3.40}) for $k=2$ can be reduced to
\begin{equation}
u^{i}_{2}\!\!\bigm\vert_{r=1}= - u^{a}_{1}\!\!\bigm\vert_{\xi=0}+
\pr_{\theta}\left( \sin\theta \, f u^{i}_{0}\!\!\bigm\vert_{r=1}\right)=
- u^{a}_{1}\!\!\bigm\vert_{\xi=0} +\cos 2\theta \, f f^{\prime}. \label{4.32}
\end{equation}
The oscillatory part of (\ref{4.32}) gives us $a_{2}(\theta,\tau)$:
\begin{equation}
a_{2}(\theta,\tau)=
- \tilde{u}^{a}_{1}\!\!\bigm\vert_{\xi=0} +\cos 2\theta \, \widetilde{f f^{\prime}}. \label{4.33}
\end{equation}
The second term on the right side of (\ref{4.33}) will be ignored because
it makes zero contribution to all quantities
which we are interested in.
The oscillatory part of (\ref{3.42}) yields $b_{2}(\theta,\tau)$:
\begin{equation}
b_{2}(\theta,\tau)=
- \tilde{u}^{b}_{1}\!\!\bigm\vert_{\eta=0}. \label{4.34}
\end{equation}
Substituting (\ref{4.27}) in (\ref{4.33}) and (\ref{4.30}) in (\ref{4.34}) (and ignoring
the second term on the right side of (\ref{4.33}) as well as the first term in (\ref{4.27})),
we solve problem (\ref{3.15}). The solution is
given by
\begin{equation}
\phi_{2}=- \frac{f\cos\theta}{(R^2-1)(R-1)^2}\left((R+1)^2 r+\frac{R^4-2R^3+6R^2-2R+1}{r}\right). \label{4.35}
\end{equation}
{\it Inner cylinder.}
Consider now Eqs. (\ref{3.24})--(\ref{3.26})
for $k=2$.
First we note that it follows from Eqs. (\ref{4.2}) and (\ref{4.6}) that
$u^{a}_{0\tau}=\nu u^{a}_{0\xi\xi}$,
which, in turn, implies that $G_{2}^{a}\equiv 0$ (see Eq. (\ref{A5}) in Appendix A). Hence, Eq. (\ref{3.25}) reduces to
$p^{a}_{2\xi}=0$.
This equation and the condition of decay at infinity
imply that $p^{b}_{2}\equiv 0$. Equation (\ref{3.24}) takes the form
\begin{equation}
v^{a}_{2\tau}-\nu v^{a}_{2\xi\xi}=F_{2}^{a}. \label{4.37}
\end{equation}
Averaging yields the equation
$\bar{v}^{a}_{2\xi\xi}=-\frac{1}{\nu}\bar{F}_{2}^{a}$
where $\bar{F}_{2}^{a}(\xi,\theta)$ is obtained by averaging Eq. (\ref{A2}).
The solution of this equation that satisfies the condition of decay at infinity can be written as
\begin{equation}
\bar{v}^{a}_{2}=-\frac{1}{\nu}\int\limits_{\xi}^{\infty} \int\limits_{\xi'}^{\infty}
\bar{F}_{2}^{a}(\xi'',\theta) \, d\xi'' \, d\xi'. \label{4.39}
\end{equation}
Lengthy, but standard calculations result in the formula
\begin{equation}
\bar{v}^{a}_{2} =\frac{R \, e^{-\tilde{\xi}}\sin(2\theta)}{4(R^2-1)^2}
\left[2R^3 e^{-\tilde{\xi}}-\left(5R^3-R\right)\tilde{\xi}
\left(\cos\tilde{\xi}-\sin\tilde{\xi}\right)
+4R^3\cos\tilde{\xi}
+\left(11R^{3}+R-4\right)\sin\tilde{\xi} \right]  \quad\quad \label{4.40}
\end{equation}
where $\tilde{\xi}=\xi/\sqrt{2\nu}$. The oscillatory part of $v^{a}_{2}$ can be obtained
by separating the oscillatory part of
Eq. (\ref{4.37}) and solving it. We will not do it as we are only interested
in the steady part of the solution.
Equation (\ref{3.26}) for $k=2$ is used to obtain $\bar{u}^{a}_{2}$:
\begin{equation}
\bar{u}^{a}_{2} = \frac{\sqrt{2\nu} \, R \, e^{-\tilde{\xi}}\cos 2\theta}{4(R^2-1)^2}
\left[2R^3 e^{-\tilde{\xi}}+\left(14R^3-6R\right)\tilde{\xi}\sin\tilde{\xi}
+\left(12R^3-4\right)\sin\tilde{\xi}+
\left(20R^3-4\right)\cos\tilde{\xi}\right]\cos 2\theta . \label{4.41}
\end{equation}
{\it Outer cylinder.} Similar calculations result in the following formulae for $\bar{v}^{b}_{2}$ and $\bar{u}^{b}_{2}$:
\begin{eqnarray}
\bar{v}^{b}_{2} &=&\frac{e^{-\tilde{\eta}}\sin 2\theta}{2R(R^2-1)^2}
\left[e^{-\tilde{\eta}}-2\tilde{\eta}\left(\cos\tilde{\eta}-\sin\tilde{\eta}\right)
+ 8\sin\tilde{\eta}+2\cos\tilde{\eta}\right],  \label{4.42} \\
\bar{u}^{b}_{2} &=& \frac{\sqrt{2\nu} \, e^{-\tilde{\eta}}\cos 2\theta}{2R^2(R^2-1)^2}
\left[e^{-\tilde{\eta}}+4\tilde{\eta}\sin\tilde{\eta}  +8\sin\tilde{\eta}+
12\cos\tilde{\eta}\right] , \quad \label{4.43}
\end{eqnarray}
where $\tilde{\eta}=\eta/\sqrt{2\nu}$.

\vskip 3mm
\noindent
{\it Averaged outer flow.} The steady part of the second-order outer flow is determined from the Stokes equations
(\ref{3.16}). Averaging (\ref{4.32}) and using yields (\ref{4.26}), we obtain
\begin{equation}
\bar{u}^{i}_{2}\!\!\bigm\vert_{r=1}=0 . \label{4.44}
\end{equation}
Boundary condition for $v^{i}_{2}$ is obtained by averaging (\ref{3.41}). It can be shown that it reduces to
\begin{equation}
\bar{v}^{i}_{2}\!\!\bigm\vert_{r=1}= -\frac{3}{2} \frac{R^4}{(R^2-1)^2}\sin 2\theta . \label{4.46}
\end{equation}
Similarly, it can be shown that
\begin{eqnarray}
\bar{u}^{i}_{2}\!\!\bigm\vert_{r=R}&=&- \bar{u}^{b}_{1}\!\!\bigm\vert_{\eta=0}= 0, \label{4.47} \\
\bar{v}^{i}_{2}\!\!\bigm\vert_{r=R} &=& -\bar{v}^{b}_{2}\!\!\bigm\vert_{\eta=0}=
-\frac{3}{2} \frac{1}{R(R^2-1)^2}\sin 2\theta. \label{4.48}
\end{eqnarray}
It is convenient to introduce stream function $\bar{\psi}^{i}_{2}$ such that
$\bar{u}^{i}_{2}=\frac{1}{r}\bar{\psi}^{i}_{2\theta}$ and
$\bar{v}^{i}_{2}=-\bar{\psi}^{i}_{2 r}$.
Then the Stokes equations (\ref{3.16}) and the boundary conditions
(\ref{4.44})--(\ref{4.48}) take the form
\begin{eqnarray}
&&\nabla^{4}\bar{\psi}^{i}_{2}=0, \quad
\bar{\psi}^{i}_{2}\!\!\bigm\vert_{r=1}=\bar{\psi}^{i}_{2}\!\!\bigm\vert_{r=R}=0,  \nonumber \\
&&\bar{\psi}^{i}_{2r}\!\!\bigm\vert_{r=1}=\frac{3}{2} \frac{R^4}{(R^2-1)^2}\sin 2\theta,
\quad
\bar{\psi}^{i}_{2r}\!\!\bigm\vert_{r=R}=
\frac{3}{2} \frac{1}{R(R^2-1)^2}\sin 2\theta. \quad \label{4.49}
\end{eqnarray}
The solution of (\ref{4.49}) is given by
\begin{equation}
\bar{\psi}^{i}_{2}=\frac{3}{4(R^2-1)^4}\left(C_1+\frac{C_2}{r^2}+C_3 r^2+C_4 r^4\right)\sin 2\theta \label{4.50}
\end{equation}
where $C_{k}$ ($k=1,\dots,4$) depend only on $R$ and are given in Appendix A (Eq. (\ref{A25})).
Equation (\ref{4.50}) is in agreement with earlier results of Duck and Smith (Eq. (3.33) in
\cite{Duck1979}) and Haddon and Riley (Eq. (3.1) in
\cite{HaddonRiley1979}).
Formulae for $\bar{u}^{i}_{2}$ and $\bar{v}^{i}_{2}$ can be easily obtained from Eq. (\ref{4.50}).


\subsection{Third order equations}

\noindent
Here we are interested only in the averaged outer flow. However,
to find it, we need boundary conditions which come from the averaged boundary layers.

\vskip 3mm
\noindent
{\it Inner cylinder.}
Averaging Eqs. (\ref{3.24}) and (\ref{3.25}) for $k=3$, we get
\begin{equation}
\bar{v}^{a}_{3\xi\xi}=\frac{1}{\nu}\bar{p}^{a}_{3\theta}-\frac{1}{\nu}\bar{F}^{a}_{3} \quad {\rm and} \quad
\bar{p}^{a}_{3\xi}=\bar{G}^{a}_{3}, \label{4.53}
\end{equation}
where $\bar{F}^{a}_{3}$ and $\bar{G}^{a}_{3}$ are obtained by averaging Eqs. (\ref{A3}) and (\ref{A6}).
We do not need explicit solutions of Eqs. (\ref{4.53}), all we need
is $\bar{v}_{3}^{a}\!\!\bigm\vert_{\xi=0}$. First, we integrate the second equation (\ref{4.53}). Then
we insert $\bar{p}^{a}_{3}$ into the first equation (\ref{4.53}) and integrate it twice.
This yields
\begin{equation}
\bar{v}^{a}_{3}=-\frac{1}{\nu}\int\limits_{\xi}^{\infty}\int\limits_{\xi'}^{\infty}
\int\limits_{\xi''}^{\infty}
\bar{G}^{a}_{3\theta}(\xi''',\theta) \, d\xi''' \, d\xi'' \, d\xi'
-\frac{1}{\nu}
\int\limits_{\xi}^{\infty}\int\limits_{\xi'}^{\infty}\bar{F}^{a}_{3}(\xi'',\theta) \, d\xi'' \, d\xi'. \label{4.55}
\end{equation}
{\it Outer cylinder.} Similarly, it can be shown that
\begin{equation}
\bar{v}^{b}_{3}=-\frac{1}{\nu}\frac{1}{R}\int\limits_{\eta}^{\infty}\int\limits_{\eta'}^{\infty}
\int\limits_{\eta''}^{\infty}
\bar{G}^{b}_{3\theta}(\eta''',\theta) \, d\eta''' \, d\eta'' \, d\eta'
-\frac{1}{\nu}
\int\limits_{\eta}^{\infty}\int\limits_{\eta'}^{\infty}\bar{F}^{b}_{3}(\eta'',\theta) \, d\eta'' \, d\eta'. \label{4.56}
\end{equation}
Here $\bar{F}^{b}_{3}$ and $\bar{G}^{b}_{3}$ are obtained by averaging Eqs. (\ref{A12}) and (\ref{A15}).

\vskip 3mm
\noindent
{\it Averaged outer flow.} The steady part of the third-order outer flow is a solution of
(\ref{3.19}) subject to appropriate boundary conditions. These boundary conditions are obtained by
averaging (\ref{3.40})--(\ref{3.43}) for $k=3$. Averaging Eqs.  (\ref{3.40}) and (\ref{3.42}) yields
the equations $\bar{u}^{i}_{3}\!\!\bigm\vert_{r=1}=- \bar{u}^{a}_{2}\!\!\bigm\vert_{\xi=0}$ and
$\bar{u}^{i}_{3}\!\!\bigm\vert_{r=R}=- \bar{u}^{b}_{2}\!\!\bigm\vert_{\eta=0}$.
Substituting (\ref{4.41}) and (\ref{4.43}) into these equations, we obtain
\begin{equation}
\bar{u}^{i}_{3}\!\!\bigm\vert_{r=1} =
-\frac{\sqrt{2\nu} \, R(11R^3-2)}{2(R^2-1)^2}\cos 2\theta, \quad
\bar{u}^{i}_{3}\!\!\bigm\vert_{r=R} =
\frac{13\sqrt{2\nu}}{2R^2(R^2-1)^2}\cos 2\theta. \label{4.57}
\end{equation}
On averaging Eq.  (\ref{3.41}) for $k=3$ and using zeroth- and first-order boundary conditions,
the boundary condition for $\bar{v}^{i}_{3}$ at $r=1$ can be simplified to
\[
\bar{v}^{i}_{3}\!\!\bigm\vert_{r=1} = - \bar{v}^{a}_{3}\!\!\bigm\vert_{\xi=0}
- \cos\theta \, \overline{f\left( v^{i}_{1r}\!\!\bigm\vert_{r=1}
+v^{a}_{2\xi}\!\!\bigm\vert_{\xi=0}\right)}
- \cos^2\theta \overline{\frac{f^2}{2}v^{a}_{1\xi\xi}\!\!\bigm\vert_{\xi=0}}
- \cos^3\theta \overline{\frac{f^3}{6}v^{a}_{0\xi\xi\xi}\!\!\bigm\vert_{\xi=0}}. \nonumber
\]
Substituting here Eq. (\ref{4.55}) and the explicit formulae for
$\bv^{i}_{0}$, $\bv^{i}_{1}$, $u^{a}_{0}$, $u^{a}_{1}$, $v^{a}_{0}$, $v^{a}_{1}$, etc.,
we find that
\begin{equation}
\bar{v}^{i}_{3}\!\!\bigm\vert_{r=1} =
\sqrt{2\nu} \, K_{1}(R)  \sin 2\theta , \quad
K_{1}(R)=\frac{R \left(53R^4-53R^3-12R^2+4R-4\right)}{4(R-1)(R^2-1)^2} . \label{4.59}
\end{equation}
Similarly, by averaging (\ref{3.43}) and using (\ref{4.56}), we obtain
\begin{equation}
\bar{v}^{i}_{3}\!\!\bigm\vert_{r=R}=-\sqrt{2\nu} \, K_{2}(R) \sin 2\theta , \quad
K_{2}(R)= \frac{4R^5+R^4-R^3+14R-14}{4R^2(R-1)(R^2-1)^2} . \label{4.60}
\end{equation}
On introducing stream function $\bar{\psi}^{i}_{3}$ such that
$\bar{u}^{i}_{3}=\frac{1}{r}\bar{\psi}^{i}_{3\theta}$ and
$\bar{v}^{i}_{3}=-\bar{\psi}^{i}_{3 r}$, Eqs. (\ref{3.19}) and boundary conditions
(\ref{4.57})--(\ref{4.60}) can be written as
\begin{eqnarray}
&&\nabla^{4}\bar{\psi}^{i}_{3}=0, \nonumber \\
&&\bar{\psi}^{i}_{3}\!\!\bigm\vert_{r=1}=-\sqrt{2\nu} \, \frac{R(11R^3-2)}{4(R^2-1)^2}\sin 2\theta,  \quad
\bar{\psi}^{i}_{3}\!\!\bigm\vert_{r=R}= \sqrt{2\nu} \, \frac{13}{4R^2(R^2-1)^2}\sin 2\theta,  \nonumber \\
&&\bar{\psi}^{i}_{3r}\!\!\bigm\vert_{r=1}= -\sqrt{2\nu} \, K_{1}(R) \sin 2\theta,  \quad
 \bar{\psi}^{i}_{3r}\!\!\bigm\vert_{r=R}=\sqrt{2\nu} \, K_{2}(R) \sin 2\theta . \label{4.61}
\end{eqnarray}
The solution of this boundary value problem is given by
\begin{equation}
\bar{\psi}^{i}_{3}=\frac{\sqrt{2\nu}}{8R(R-1)(R^2-1)^4}
\left(D_1+\frac{D_2}{r^2}+D_3 r^2+D_4 r^4\right)\sin 2\theta \label{4.62}
\end{equation}
where $D_{k}$ ($k=1,\dots,4$) depend on $R$ only and are given by Eq. (\ref{A26}) in Appendix A.
Explicit expressions for $\bar{u}^{i}_{3}$ and $\bar{v}^{i}_{3}$ can now be easily obtained from (\ref{4.62}).


\subsection{Stokes drift}

\noindent
So far we have discussed the Eulerian velocity. However, the velocity observed in experiments is the velocity
of fluid particles, i.e. the Largangian velocity. It is well-known that in oscillatory flows the observed
averaged Lagrangian velocity differs from the averaged Eulerian velocity, and the difference between these two is
known as the Stokes drift.
Below we discuss the effect of the Stokes drift on the averaged Largangian velocity.

\vskip 3mm
\noindent
The motion of fluid particles is governed by the ordinary differential
equation
\begin{equation}
\frac{d\bx}{d\tau}=\epsilon^2\bv(\bx,\tau,\eps), \label{4.65}
\end{equation}
The velocity field $\bv(\bx,\tau,\eps)$ is the solution of the Navier-Stokes equations, which is
$2\pi$-periodic in $\tau$ and has a nonzero average.
We have already computed first three terms in the uniformly valid asymptotic expansion of $\bv(\bx,\tau,\eps)$.
Now we are interested in constructing an asymptotic expansion of the solution
of (\ref{4.65}) for small $\eps$. We introduce the slow time $t=\eps^2\tau$, assume that $\bx=\bx(\tau,t,\eps)$
and substitute this in (\ref{4.65}). This gives us the equation
\[
\bx_{\tau}=\epsilon^2\left(\bv(\bx,\tau,\eps)-\bx_{t}\right).
\]
It is shown in Appendix $B$ that the solution of this equation
can be presented in the form
$\bx(x_{0},\tau)=\tilde{\bx}(x_{0},\tau, t, \eps)+ \bar{\bx}(x_{0}, t, \eps)$ where
$\tilde{\bx}(\bx_{0},\tau, t, \eps)$ represents purely oscillatory
part of the motion of the fluid particle whose averaged position at $t=0$ was $\bx_{0}$ and
$\bar{\bx}(x_{0}, t, \eps)$, is the solution of the equation
\[
\bx_{t}=\bar{\bv}(\bx,\eps)+\bar{\bv}^{s}(\bx,\eps).
\]
this equation describes slow motion of this particle due to the steady part of the Eulerian velocity
field, $\bar{\bv}(\bx,\eps)$, and the Stokes drift velocity, $\bar{\bv}^{s}(\bx,\eps)$.
If we denote
the Lagrangian velocity of fluid particles by superscript $L$ and the Eulerian velocity by superscript $E$,
then our results can be summarized
as follows. Our asymptotic expansion for the averaged Eulerian velocity has the form
\begin{equation}
\bar{u}^{E}= \, \eps^2\left[\bar{u}_{2}^{i}+\bar{u}_{1}^{a}\right]+ \, O(\eps^{3}), \quad
\bar{v}^{E}= \, \eps \, \bar{v}_{1}^{a}
+ \, \eps^2\left[\bar{v}_{2}^{i}+\bar{v}_{2}^{a}+\bar{v}_{2}^{b}\right]+ \, O(\eps^{3}). \label{4.66}
\end{equation}
It is shown in Appendix B that
the Lagrangian velocity of fluid particles is given by
\begin{equation}
\bar{u}^{L}= \, \eps^2 \bar{u}_{2}^{i} + \, O(\eps^{3}), \quad
\bar{v}^{L}= \, \eps^2 \left[\bar{v}_{2}^{i}+\bar{v}_{2}^{a}+\bar{v}_{2}^{b}
+\bar{v}_{2}^{s}\right]+ \, O(\eps^{3}). \label{4.68}
\end{equation}
where the Stokes drift velocity of the fluid particles $\bar{v}_{2}^{s}$ is given by Eq. (\ref{5.46})
in Appendix B.
Comparing (\ref{4.66}) with (\ref{4.68}), we observe that
the Stokes drift eliminates (i) $\bar{u}_{1}^{a}$ from the first equation (\ref{4.66})
and (ii) the $O(\eps)$ term from the second equation (\ref{4.66}). It also results in
the additional $O(\eps^2)$ term
$\bar{v}_{2}^{s}$ in the expansion of the azimuthal velocity. Thus, {\em the $O(\eps)$ steady boundary layer at
the inner cylinder
disappears when we take account of the Stokes drift}. This is a consequence of the fact that
the steady Lagrangian velocity rather than the steady Eulerian velocity are invariant to the change of
reference frame. More precisely, the steady Eulerian velocity is not invariant in the following sense:
it can be shown that in the reference frame fixed in the inner cylinder, there would
be no $O(\eps)$ steady boundary layer at the inner cylinder and an $O(\eps)$ steady boundary layer
would appear near the outer cylinder (which is oscillating in this reference frame). It can be also shown that
the steady Lagrangian velocity is the same both in the oscillating reference frame and in the fixed one.
We note here that if one uses the method of matched asymptotic expansions, then although it is possible,
it is not easy to detect the existence of an
$O(\eps)$ steady boundary layer near the inner or outer cylinders. Fortunately,
the presence or absence of the $O(\eps)$ steady boundary layer do not affect the outer flow:
the $O(\eps)$ steady outer flow is zero in both cases.

\vskip 3mm
\noindent
Further calculations with the help of the known formulae for
$\bar{v}_{2}^{a}$, $\bar{v}_{2}^{b}$ and $\bar{v}_{2}^{s}$ show that
$\bar{v}^{L}$ can be written as
\begin{equation}
\bar{v}^{L}=  \eps^2 \left[\bar{v}_{2}^{i}+(\bar{v}_{2}^{a})^{L}
+(\bar{v}_{2}^{b})^{L}\right] + \, O(\eps^{3}),  \label{4.70}
\end{equation}
where
\begin{equation}
\bigl(\bar{v}_{2}^{a}\bigr)^{L} =-\frac{R^4 \sin 2\theta}{4(R^2-1)^2} \,
F'(\tilde{\xi}) ,  \quad
\bigl(\bar{v}_{2}^{b}\bigr)^{L} =-\frac{\sin 2\theta}
{4R(R^2-1)^2} \,
F'(\tilde{\eta}),   \label{4.71}
\end{equation}
where $F(z)=e^{-z}\left(3e^{-z}+8\cos z+8\sin z\right)$, $\tilde{\xi}=\xi/\sqrt{2\nu}$
and $\tilde{\eta}=\eta/\sqrt{2\nu}$.

\subsection{Asymptotic expansion for stream function}

\noindent
To rewrite our asymptotic expansion of terms of the averaged stream function, we first observe that
\[
\psi=\psi_{0}^{i}+\eps\left[\psi_{1}^{i}+\psi_{0}^{a}
+\psi_{0}^{b}\right]+\eps^2\left[\psi_{2}^{i}+\psi_{1}^{a}
+\psi_{1}^{b}\right]+\eps^3\left[\psi_{3}^{i}+\psi_{2}^{a}+\psi_{2}^{b}\right]+O(\eps^4),
\]
where $\psi_{k}^{i}$ is such that $u_{k}^{i}=\frac{1}{r}\psi_{k\theta}^{i}$ and
$v_{k}^{i}=-\psi_{k r}^{i}$ for $k=0,1,\dots$ and where $\psi_{k}^{a}$, $\psi_{k}^{b}$ are defined as
\[
\psi_{k}^{a}=\int\limits_{\xi}^{\infty}v_{k}^{a}(\xi',\theta,\tau)d\xi', \quad
\psi_{k}^{b}= -\int\limits_{\eta}^{\infty}v_{k}^{b}(\eta',\theta,\tau)d\eta'
\]
for $k=0,1,\dots$
Similarly, we have
$\bar{\psi}=\eps^2\left[\bar{\psi}_{2}^{i}+\bar{\psi}_{1}^{a}
\right]+\eps^3\left[\bar{\psi}_{3}^{i}+\bar{\psi}_{2}^{a}+\bar{\psi}_{2}^{b}\right]+O(\eps^4)$
and
\begin{equation}
\bar{\psi}^{L}=\eps^2\bar{\psi}_{2}^{i}
+\eps^3\left[(\bar{\psi}_{3}^{i})^{L}+(\bar{\psi}_{2}^{a})^{L}+(\bar{\psi}_{2}^{b})^{L}\right]+O(\eps^4). \label{4.73}
\end{equation}
In the last formula, $(\bar{\psi}_{2}^{a})^{L}$ and
$(\bar{\psi}_{2}^{b})^{L}$  are obtained from (\ref{4.71}) and given by
\begin{eqnarray}
\bigl(\bar{\psi}_{2}^{a}\bigr)^{L} =\frac{\sqrt{2\nu} \, R^4e^{-\frac{\xi}{\sqrt{2\nu}}}}{4(R^2-1)^2}
\, F(\tilde{\xi})\sin 2\theta, \quad \quad
\bigl(\bar{\psi}_{2}^{b}\bigr)^{L} =-\frac{\sqrt{2\nu}}{4R(R^2-1)^2}
\, F(\tilde{\eta})\sin 2\theta.  \label{4.74}
\end{eqnarray}
$(\bar{\psi}_{3}^{i})^{L}$ is the stream function for the third-order Lagrangian velocity (see Appendix B):
\begin{equation}
\left(\bar{\psi}^{i}_{3}\right)^{L}=\frac{\sqrt{2\nu}}{8R(R-1)(R^2-1)^4}
\left(D_1+\frac{\hat{D}_2}{r^2}+D_3 r^2+D_4 r^4\right)\sin 2\theta, \label{4.76}
\end{equation}
where $D_1$, $\hat{D}_2$, $D_3$ and $D_4$ are constants (given by (\ref{A26}) and (\ref{BBB})).

\vskip 3mm
\noindent
Thus, the Stokes drift produces a nonzero contribution to the third order outer flow given by the second terms
on the right sides of Eqs. (\ref{5.58}) and (\ref{5.59}). Note that the Stokes drift
corrections to the outer flow in the lower order approximations are all zero.
Also, it follows from Eqs. (\ref{5.58}) and (\ref{5.59}) that the Stokes drift contribution to
$\left(\bar{\psi}^{i}_{3}\right)^{L}$ vanishes in the limit $R\to\infty$, which means that
if there were no outer cylinder, the third order Stokes drift
correction to the outer flow would be zero too. This can
also be shown independently by treating the problem on a steady flow produced
by an oscillating cylinder in an unbounded fluid.
Thus, the appearance of a nonzero Stokes drift
correction to the outer flow is caused by the presence of the outer cylinder.


\setcounter{equation}{0}
\renewcommand{\theequation}{5.\arabic{equation}}

\section{Discussion}


\vskip 3mm
\noindent
Equation (\ref{4.73}) together with Eqs. (\ref{4.50}), (\ref{4.74})--(\ref{4.76}) represent the first
two non-zero terms
in the asymptotic expansion of the stream function for the averaged Lagrangian velocity.
Let us first discuss the domain of applicability of formula
(\ref{4.73}).

\vskip 3mm
\noindent
{\em Domain of applicability}.
Our asymptotic expansion is formally valid for $\eps \ll 1$ and for $\nu=O(1)$. In practice, we may expect that it will
be valid for all $\eps$ and $\nu$ such that the contribution of the $O(\eps^3)$ term to the right side
of Eq. (\ref{4.73}) is smaller than the contribution of the $O(\eps^2)$ term.
For each value of $R>1$, this requirement corresponds to a domain in the space of parameters
$\eps$ and $\nu$. It is convenient to rewrite Eq. (\ref{4.73}) in the form
\[
\bar{\psi}^{L}=\eps^2\left[\Phi_{0}(R,r)+ \eps\sqrt{\nu}\Phi_{1}(R,r,\nu,\eps)\right] \sin 2\theta + O(\eps^4)
\]
where $\Phi_{0} = \bar{\psi}_{2}^{i} \, /\sin 2\theta$ and
$\Phi_{1} = \frac{1}{\sqrt{\nu}}\left[(\bar{\psi}_{3}^{i})^{L}+
(\bar{\psi}_{2}^{a})^{L}\!\bigm\vert_{\xi=(r-1)/\eps}
+(\bar{\psi}_{2}^{b})^{L}\!\bigm\vert_{\eta=(R-r)/\eps}\right]/\sin 2\theta$.

\begin{figure}
\begin{center}
\includegraphics*[trim=0mm 0mm 0mm 7mm,height=6cm]{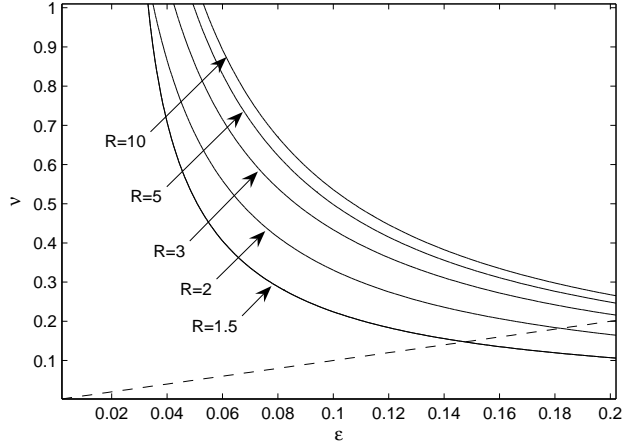}
\end{center}
\vskip -8mm
\caption{The curves $\varkappa (R,\nu,\eps)=1$ for $R=2,3,5 \ {\rm and} \ 10$. The dashed line corresponds to
$\nu=\eps$. For each $R$, the `domain of applicability' of the theory lies below the corresponding curve, but
above the dashed line.}
\label{fig1}
\end{figure}

\vskip 3mm
\noindent
Consider now the following quantity
\[
\varkappa (R,\nu,\eps) = \eps\sqrt{\nu}\max_{r\in[1,R]}\left\vert\Phi_{1}(R,r,\nu,\eps)\right\vert
/ \max_{r\in[1,R]}\left\vert\Phi_{0}(R,r)\right\vert,
\]
which measures the magnitude of the second nonzero term relative to the first term. We expect that
our theory will work for all $R$, $\nu$, $\eps$ such that $\varkappa (R,\nu,\eps)< 1$, and the smaller
$\varkappa$ is, the better the theory should work. The level curves $\varkappa (R,\nu,\eps)=1$ in the $(\eps,\nu)$
plane for several values of $R$ are shown in Fig. \ref{fig1}. For each $R$ the domain, where we expect
that the theory is valid,
lies below the corresponding curve. Of course, the `domains of applicability' shown in Fig. \ref{fig1}
are very approximate.
Nevertheless Fig. \ref{fig1} gives us an idea of where our theory might work.
In particular, it shows that for $\nu\geq 1$ the interval in $\epsilon$ within which the theory is applicable
is very narrow, but it becomes much wider for smaller values of $\nu$. Figure \ref{fig1} also shows that the `domain of
applicability' shrinks when we reduce
the radius $R$ of the outer cylinder. Our theory is also unapplicable to the case of high Reynolds numbers when the
dimensionless viscosity $\nu$ is comparable with $\eps$. The dashed-line curve in Fig. \ref{fig1} corresponds to
$\nu/\eps=1$, and the area below this curve is not covered by the present theory.

\begin{figure}
\begin{center}
\includegraphics*[height=8cm]{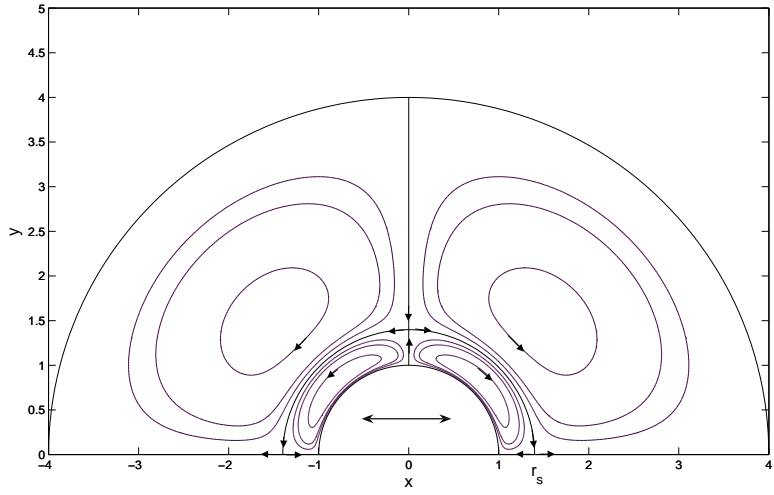}
\end{center}
\vskip -10mm
\caption{Typical streamlines of the averaged Lagrangian flow.}
\label{fig2}
\end{figure}

\begin{figure}
\begin{center}
\includegraphics*[height=8cm]{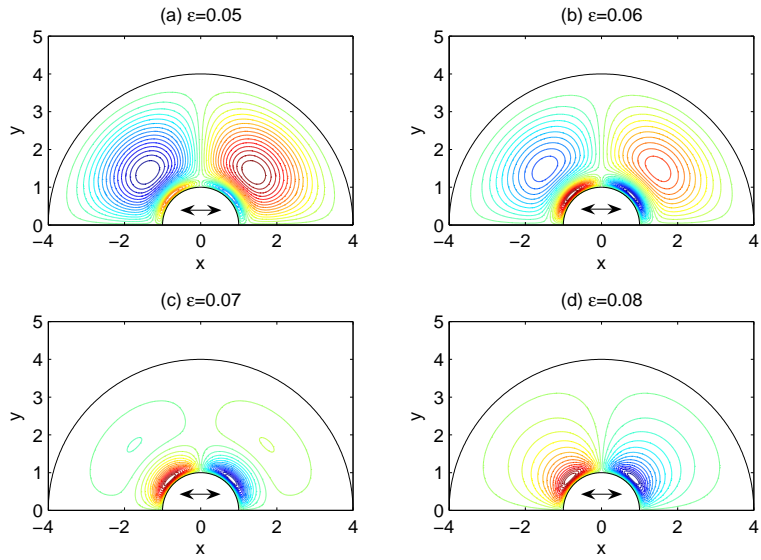}
\end{center}
\vskip -10mm
\caption{The streamlines of the averaged Lagrangian flow for $\nu=0.5$, $R=4$}
\label{fig3}
\end{figure}

\vskip 3mm
\noindent
{\em Comparison with experiments}. Schematic picture of the averaged Lagrangian flow is shown in Fig. 2.
The flow is symmetric relative to the axis of oscillation (the $x$-axis).
There are four stagnation points in the flow (two on
the $x$ axis and two on the $y$ axis) and three of
them are shown in Fig. \ref{fig2}. The stagnation points are symmetric relative to the $x$ and $y$ axes
and their distance from the origin is denoted by $r_{s}$ in Fig. \ref{fig2} (note that
the thickness of the inner vortex structure is $r_{s}-1$).
Examples of the streamlines produced by formula (\ref{4.73}) are shown in Fig. \ref{fig3}.
The streamlines shown in Figures \ref{fig3}a, \ref{fig3}b and \ref{fig3}c
(which correspond to $\eps=0.05$, $0.06$ and $0.07$
for $\nu=0.5$ and $R=4$) are similar and qualitatively consistent with the experimental observations
(see, e.g., \cite{Bertelsen1973}, \cite{Tatsuno}).
Figures \ref{fig3}a, \ref{fig3}b and \ref{fig3}c show that the thickness and the intensity of
the inner vortex structure increases
with $\eps$. At some value of $\eps$ the inner vortex structure becomes dominating (Fig. \ref{fig3}d). This
is inconsistent with the experimental observations and corresponds to the situation where
the $O(\eps^3)$ term in (\ref{4.73}) is larger than (or comparable to) the  $O(\eps^2)$ term.
Our theory is not valid in this situation.

\begin{figure}
\begin{center}
\includegraphics*[height=6cm]{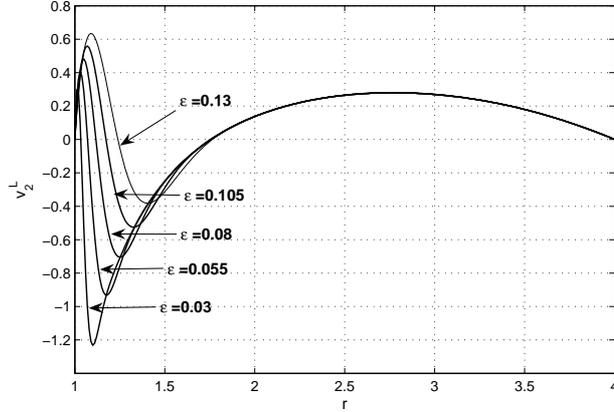}
\vskip -5mm
\end{center}
\vskip -10mm
\caption{The steady azimuthal velocity $\bar{v}^{L}=\bar{v}_{2}^{i}+(\bar{v}_{2}^{a})^{L}+(\bar{v}_{2}^{b})^{L}$
as a function of $r$ for $\theta=\pi/4$, $\nu=0.5$ and $R=4$}
\label{fig4}
\end{figure}

\vskip 3mm
\noindent
Typical profiles of the azimuthal velocity are shown in Fig. \ref{fig4}.
Qualitatively the behaviour of the azimuthal velocity
is similar to the velocity profiles measured in the experiments by Bertelsen et. al. \cite{Bertelsen1973}, but there
is no quantitative agreement, because in the experiments $\eps\approx 0.3$, which is outside
the area of applicability of
our theory.

\begin{figure}
\begin{center}
\includegraphics*[height=6cm]{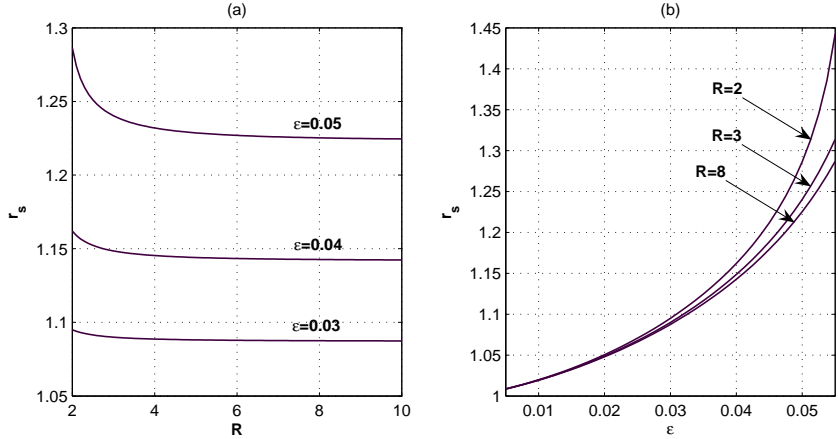}
\end{center}
\vskip -8mm
\caption{(a) The distance of the stagnation points from the origin $r_{s}$ versus $R$; (b) $r_{s}$ versus $\eps$.}
\label{fig5}
\end{figure}

\begin{figure}
\begin{center}
\includegraphics*[height=7cm]{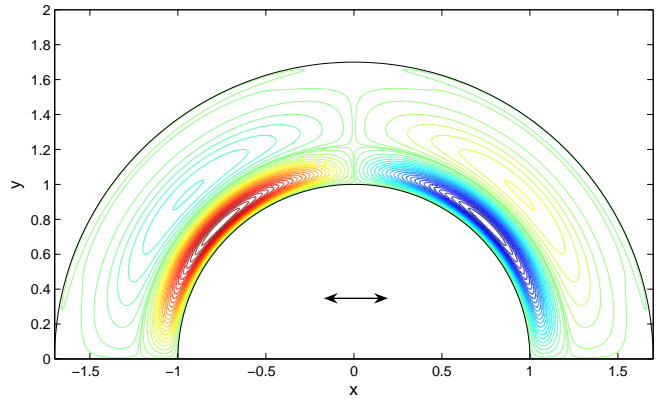}
\vskip -5mm
\end{center}
\vskip -15mm
\caption{The streamlines of the averaged flow for $\nu=0.4$, $\eps=0.048$ and $R=1.7$}
\label{fig6}
\end{figure}

\vskip 3mm
\noindent
The averaged flow structure does not change for a wide range of values of $R$. Figure \ref{fig5}
shows the dependence of the distance of stagnation points from the origin $r_{s}$ on the radius of the outer cylinder
$R$  and on $\eps$. Qualitatively the behaviour of $r_{s}$ (as a function of $R$)
agrees with experimental observation of Bertelsen et. al. \cite{Bertelsen1973}.
The structure of the averaged flow changes slightly for smaller values of $R$ (when $R$ is not very different from 1).
In this case, the boundary layer at the outer cylinder results in appearance of an additional
narrow vortex system near the outer cylinder. Figure \ref{fig6} shows the streamlines
for the Lagrangian velocity for $\nu=0.4$, $\eps=0.048$ and $R=1.7$. One can see a narrow boundary layer at the
outer cylinder which is similar to the boundary layer at the inner cylinder but considerably weaker.

\vskip 3mm
\noindent
{\em One cylinder in the fluid that extends to infinity}.
In the limit $R\to\infty$, the stream function
for the Lagrangian velocity
reduces to
\begin{equation}
\bar{\psi}^{L}=\eps^2 \bar{\psi}^{i}_{2} +
\eps^3 \left[ \bar{\psi}^{i}_{3}+\left(\bar{\psi}_{2}^{a}\right)^{L}\right]+O(\eps^4), \label{6.3}
\end{equation}
where
\[
\bar{\psi}^{i}_{2} =\frac{3}{4}\left(1-\frac{1}{r^2}\right)\sin 2\theta,  \quad
\bar{\psi}^{i}_{3} =\frac{\sqrt{2\nu}}{8}
\left(-75+\frac{53}{r^2}\right)\sin 2\theta, \quad
\left(\bar{\psi}_{2}^{a}\right)^{L} =\frac{\sqrt{2\nu}}{4}\, F(\tilde{\xi})\, \sin 2\theta .
\]

\noindent
Wang \cite{Wang1968} studied the steady flow produced be a fixed circular cylinder placed in an oscillating
flow. He applied the method of matched asymptotic expansions and obtained a uniformly valid expansion of the
stream function under the same assumptions as in the present paper. His theory however was not directly applicable
to the steady streaming flow produced by a cylinder oscillating in the fluid which is at rest at infinity,
because his expansion of the Eulerian velocity field is not invariant under the appropriate
change of the reference frame. In order to obtain the invariant velocity filed,
one should consider the Lagrangian velocity which is different from the Eulerian velocity
by the Stokes drift velocity of the fluid particles. This had been understood first
by Skavlem and Tj$\o$tta \cite{Skavlem1955}, and Bertelsen et. al. \cite{Bertelsen1973}
had corrected Wang's theory by taking account of the Stokes drift. However,
the averaged stream function for the Eulerian velocity in Wang's theory
(given by Eq. (3.36) in \cite{Wang1968} and Eq. (25) in \cite{Bertelsen1973})
is incomplete because of the absence of the $O(\eps^3)$ term associated with the averaged
outer flow. In order to compute this term, it is necessary to obtain first the second
nonzero term in the inner expansion of the averaged stream function and then, following Van Dyke's
recipe \cite{Van Dyke1964},
match the first two terms of the outer expansion with
the first two terms of the inner expansion. This will result in a correct $O(\eps^3)$ outer flow term, which
then can be incorporated in the composite formula valid for the whole flow domain.
It can be shown that if this is done, then the uniformly valid expansion for the averaged stream function
becomes
\[
\bar{\psi}=\eps^2 \biggl\{\frac{3}{4}\left(1-\frac{1}{r^2}\right)+
\eps \, \frac{\sqrt{2\nu}}{4}\biggl[ \frac{1}{2}
\left(-75+\frac{49}{r^2}\right)
+ G(\tilde{\xi})\biggr]\biggr\}
\sin 2\theta +O(\eps^4)\quad
\]

\noindent
where $G(\tilde{\xi})=e^{-\tilde{\xi}}
(e^{-\tilde{\xi}}
+12\cos\tilde{\xi}+8\sin\tilde{\xi}+4\tilde{\xi}
\sin\tilde{\xi})$.
Exactly the same formula can be obtained by the Vishik-Lyusternik method. If we now
take into account the Stokes drift, we obtain Eq. (\ref{6.3}). This proves that
(at least) up to $O(\eps^3)$ terms, the averaged stream function for the Lagrangian velocity
is the same both in the oscillating (with the cylinder) reference frame and
in the fixed (in space) reference frame.

\vskip 3mm
\noindent
{\em A remark on the results of Duck and Smith} \cite{Duck1979}. As was mentioned earlier,
the expression for $\bar{\psi}^{i}_{2}$ obtained here agrees with formula (3.33) in \cite{Duck1979}
(as well as with equation (3.1) in
\cite{HaddonRiley1979}).
However, the boundary layer parts of the expansions are different. The most essential difference
is that in our expansion of the Eulerian velocity there is a nonzero steady boundary layer at the inner cylinder in the
first order in $\epsilon$, while it is present only in the second order in \cite{Duck1979}.
This is a result of the transformation of coordinates employed in \cite{Duck1979}.
The transformation maps the gap between two eccentric cylinders onto the annulus between two cylinders
whose axes coincide and it is a time-dependent transformation. In order to obtain our expansion, one needs to
do the following: (i) to write down a composite expansion using the formulae for inner and outer expansions
obtained in \cite{Duck1979}, (ii) to perform the inverse transformation of the coordinates, and
(iii) to expand the resulting velocity field in Taylor's series in $\epsilon$. The result will almost certainly
coincide with our expansion of the averaged stream function for the Eulerian velocity up to $O(\eps^2)$ terms.

\vskip 3mm
\noindent
{\em Remarks of the steady streaming theories for high Reynolds numbers}.
There are quite a few papers dealing with
the steady streaming in an unbounded fluid produced by an oscillating circular cylinder (see
\cite{Stuart1963,Stuart1966,Riley1965,Riley1967,Riley1975}) at high Reynolds numbers such that
$Re_{s}=O(1)$ or $Re_{s} \gg 1$ (in our notation this means that $\nu/\eps^2=O(1)$ or $\nu/\eps^2 \ll 1$,
respectively).
In all these papers, coordinate systems oscillating with the cylinder are used, i.e. what is being really solved is
the problem about a steady streaming flow produced by the fixed cylinder placed in an unbounded oscillating flow.
As the present study shows, the problem with a fixed cylinder is not equivalent to the problem with an oscillating
cylinder not only within the boundary layer but also in the outer flow. In fact, the situation is even more difficult,
because for $\nu/\eps^2=O(1)$ (and even more so for $\nu/\eps^2 \ll 1$) it is unclear how to solve the problem
in the frame of reference fixed in space. In the present study,
we employed the standard trick: using the fact that the amplitude of oscillations was small we expanded the solution
in Taylor's series about the averaged position of the oscillating cylinder and thus transferred the no-slip
boundary conditions at the moving surface of the oscillating cylinder to the fixed
surface of the cylinder at its averaged position. This allowed us to formulate our asymptotic expansion
in terms of a sequence of boundary value problems with fixed boundaries. The thickness of the boundary layer
on the inner cylinder that appeared in our expansion is $O(\eps)$, which is much larger than
the $O(\eps^2)$ displacement of the inner cylinder from its averaged position, and this
is what justifies the transfer of boundary conditions from the moving boundary to the fixed one.
If however we considered the case of $\nu=\eps^2$ ($Re_{s}=1$), the thickness of the Stokes layer
would be $O(\eps^2)$, i.e. of the same order as the displacement of the cylinder, and therefore it would
be impossible to justify the transfer of boundary conditions. Even if we worked in the frame of reference
fixed with
the oscillating cylinder, we would encounter the problem of calculating the Stokes drift velocity in the boundary
layer because of non-analytic dependence of the boundary layer velocity on $\eps$.

\vskip 3mm
\noindent
In \cite{HaddonRiley1979} and \cite{Duck1979}, the same problem as in the present paper, i.e.
the steady flow between two cylinders produced by
small-amplitude oscillations of the inner cylinder, had been studied in the case of $Re_{s}=O(1)$
(and $Re_{s}\gg 1$). Haddon and Riley \cite{HaddonRiley1979} had realised the impossibility of
the transfer of the boundary conditions from the moving surface to a
fixed one in this flow regime and addressed the problem by employing two different coordinate systems for
boundary layers on the inner and outer cylinders. In the end, however, they
had applied the transfer of the boundary conditions at the inner cylinder for the steady outer flow,
and it is unclear whether this can be justified.
In \cite{Duck1979}, a conformal mapping that maps the gap between two eccentric cylinders
onto the  annulus between two cylinders with a common axis was employed. The subsequent analysis had been done using
the transformed coordinates. Neither the inverse transformation to physical coordinates, nor the Stokes drift
had been computed, and as the  above discussion indicates, these are the questions where potential
problems may arise. Thus, in spite of a considerable progress in this area and a good agreement with experimental
results achieved by the theory (see, e.g., \cite{Duck1979}), there are still certain unanswered questions
concerning the steady streaming at high Reynolds numbers.

\setcounter{equation}{0}
\renewcommand{\theequation}{A.\arabic{equation}}

\section{Appendix A. Explicit expressions for the right hand sides of Eqs. (\ref{3.24})--(\ref{3.26}),
(\ref{3.31})--(\ref{3.33}), (\ref{3.40}) and (\ref{3.41})}

\noindent
Functions $F^{a}_{k}$, $G^{a}_{k}$ and $H^{a}_{k}$ for $k=1,2,3$ in Eqs. (\ref{3.24})--(\ref{3.26})
are given by
\begin{eqnarray}
F^{a}_{1}&=&-u^{i}_{0}\!\!\bigm\vert_{r=1}v^{a}_{0\xi}+\nu v^{a}_{0\xi}+\xi p^{a}_{0\theta}, \label{A1} \\
F^{a}_{2}&=&-u^{i}_{0}\!\!\bigm\vert_{r=1}v^{a}_{1\xi}
-\left(u^{i}_{1}\!\!\bigm\vert_{r=1}+\xi u^{i}_{0r}\!\!\bigm\vert_{r=1}\right)v^{a}_{0\xi}
-\pr_{\theta}\left(\frac{(v^{a}_{0})^2}{2}+
v^{i}_{0}\!\!\bigm\vert_{r=1}v^{a}_{0}\right) \nonumber \\
&&- u^{a}_{0}v^{a}_{0\xi} -  \, u^{i}_{0}\!\!\bigm\vert_{r=1}v^{a}_{0}
+\nu\left(v^{a}_{1\xi}-\xi v^{a}_{0\xi}+v^{a}_{0\theta\theta}-v^{a}_{0}\right)
+\xi p^{a}_{1\theta}-\xi^2 p^{a}_{0\theta},  \label{A2} \\
F^{a}_{3}&=&-u^{i}_{0}\!\!\bigm\vert_{r=1}v^{a}_{2\xi}
-\left(u^{i}_{1}\!\!\bigm\vert_{r=1}+\xi u^{i}_{0r}\!\!\bigm\vert_{r=1}\right)v^{a}_{1\xi}
- \left(u^{i}_{2}\!\!\bigm\vert_{r=1}+\xi u^{i}_{1r}\!\!\bigm\vert_{r=1}+
\frac{\xi^2}{2} u^{i}_{0rr}\!\!\bigm\vert_{r=1}\right)v^{a}_{0\xi} \nonumber \\
&&-\pr_{\theta}\Biggl(v^{i}_{0}\!\!\bigm\vert_{r=1}v^{a}_{1}+
\left(v^{i}_{1}\!\!\bigm\vert_{r=1}+\xi v^{i}_{0r}\!\!\bigm\vert_{r=1}\right)v^{a}_{0}+
v^{a}_{0}v^{a}_{1}-\xi \frac{(v^{a}_{0})^2}{2}-\xi v^{i}_{0}\!\!\bigm\vert_{r=1}v^{a}_{0}\Biggr) \nonumber \\
&&-u^{a}_{0}v^{a}_{1\xi}-u^{a}_{1}v^{a}_{0\xi} - v^{i}_{0r}\!\!\bigm\vert_{r=1}u^{a}_{0}
- u^{i}_{0}\!\!\bigm\vert_{r=1}v^{a}_{1} -
\left( u^{i}_{1}\!\!\bigm\vert_{r=1} + \xi u^{i}_{0r}\!\!\bigm\vert_{r=1}\right)v^{a}_{0}
-v^{i}_{0}\!\!\bigm\vert_{r=1}u^{a}_{0} \nonumber \\
&&-u^{a}_{0}v^{a}_{0} - \xi u^{i}_{0}\!\!\bigm\vert_{r=1}v^{a}_{0}
+\xi p^{a}_{2\theta}-\xi^2 p^{a}_{1\theta}+\xi^3 p^{a}_{0\theta}  \nonumber \\
&& +  \nu\Bigl(v^{a}_{2\xi}-\xi v^{a}_{1\xi}+ \xi^2 v^{a}_{0\xi}+v^{a}_{1\theta\theta}
-2\xi v^{a}_{0\theta\theta}-v^{a}_{1}+2\xi v^{a}_{0}+ 2u^{a}_{0\theta}\Bigr), \quad \label{A3} \\
G^{a}_{1}&=&0,  \label{A4} \\
G^{a}_{2}&=&-\left(u^{a}_{0\tau}-\nu u^{a}_{0\xi\xi}\right),  \label{A5}  \\
G^{a}_{3}&=&-\left(u^{a}_{1\tau}-\nu u^{a}_{1\xi\xi}\right)-
u^{i}_{0}\!\!\bigm\vert_{r=1}u^{a}_{0\xi}-
u^{i}_{0\theta}\!\!\bigm\vert_{r=1}v^{a}_{0}+(v^{a}_{0})^2+2v^{i}_{0}\!\!\bigm\vert_{r=1}v^{a}_{0}
+\nu\left(u^{a}_{0\xi}-2v^{a}_{0\theta}\right)  , \quad \label{A6} \\
H^{a}_{1}&=&\xi v^{a}_{0\theta}-u^{a}_{0},  \label{A7}  \\
H^{a}_{2}&=&\xi v^{a}_{1\theta}-\xi^2 v^{a}_{0\theta}-u^{a}_{1}+\xi u^{a}_{0} ,  \label{A8} \\
H^{a}_{3}&=&\xi v^{a}_{2\theta}-\xi^2 v^{a}_{1\theta}+\xi^3 v^{a}_{0\theta}
-u^{a}_{2}+\xi u^{a}_{1}-\xi^2 u^{a}_{0} .  \label{A9}
\end{eqnarray}

\vskip 3mm
\noindent
Functions $F^{b}_{k}$, $G^{b}_{k}$ and $H^{b}_{k}$ for $k=1,2,3$ in Eqs. (\ref{3.31})--(\ref{3.33})
are given by
\begin{eqnarray}
F^{b}_{1}&=&-\frac{\nu}{R} v^{b}_{0\eta}-\frac{\eta}{R^2} p^{b}_{0\theta}, \label{A10} \\
F^{b}_{2}&=&u^{i}_{0}\!\!\bigm\vert_{r=R}v^{b}_{1\eta}
+\left(u^{i}_{1}\!\!\bigm\vert_{r=R}-\eta u^{i}_{0r}\!\!\bigm\vert_{r=R}\right)v^{b}_{0\eta}+
u^{b}_{0}v^{b}_{0\eta}-\frac{1}{R}\pr_{\theta}\left(\frac{(v^{b}_{0})^2}{2}+
v^{i}_{0}\!\!\bigm\vert_{r=R}v^{b}_{0}\right)  \nonumber \\
&&  - \, \frac{1}{R}u^{i}_{0}\!\!\bigm\vert_{r=R}v^{b}_{0} +
\nu\left(-\frac{1}{R}v^{b}_{1\eta}-\frac{\eta}{R^2} v^{b}_{0\eta}+
\frac{1}{R^2}\left(v^{b}_{0\theta\theta}-v^{b}_{0}\right)\right)
-\frac{\eta}{R^2} p^{b}_{1\theta}-\frac{\eta^2}{R^3} p^{b}_{0\theta}, \quad\quad \label{A11} \\
F^{b}_{3}&=&u^{i}_{0}\!\!\bigm\vert_{r=R}v^{b}_{2\eta}
+\left(u^{i}_{1}\!\!\bigm\vert_{r=R}-\eta u^{i}_{0r}\!\!\bigm\vert_{r=R}\right)v^{b}_{1\eta}+
\left(u^{i}_{2}\!\!\bigm\vert_{r=R}-\eta u^{i}_{1r}\!\!\bigm\vert_{r=R}+
\frac{\eta^2}{2} u^{i}_{0rr}\!\!\bigm\vert_{r=R}\right)v^{b}_{0\eta} \nonumber \\
&&-
\frac{1}{R}\pr_{\theta}\Biggl[v^{i}_{0}\!\!\bigm\vert_{r=R}v^{b}_{1}+
\left(v^{i}_{1}\!\!\bigm\vert_{r=R}-\eta v^{i}_{0r}\!\!\bigm\vert_{r=R}\right)v^{b}_{0}+
v^{b}_{0}v^{b}_{1} +\frac{\eta}{R^2}\left( \frac{(v^{b}_{0})^2}{2}+
v^{i}_{0}\!\!\bigm\vert_{r=R}v^{b}_{0}\right)\Biggr] \nonumber \\
&&-
\frac{1}{R}\left[u^{i}_{0}\!\!\bigm\vert_{r=R}v^{b}_{1} +
\left(u^{i}_{1}\!\!\bigm\vert_{r=R} - \eta u^{i}_{0r}\!\!\bigm\vert_{r=R}\right)v^{b}_{0}
+v^{i}_{0}\!\!\bigm\vert_{r=R}u^{b}_{0}
+u^{b}_{0}v^{b}_{0}\right]  \nonumber \\
&& +u^{b}_{0}v^{b}_{1\eta}
+u^{b}_{1}v^{b}_{0\eta} - v^{i}_{0r}\!\!\bigm\vert_{r=R}u^{b}_{0} -\frac{\eta}{R^2}
u^{i}_{0}\!\!\bigm\vert_{r=R}v^{b}_{0} -\frac{\eta}{R^2} p^{b}_{2\theta}-\frac{\eta^2}{R^3} p^{b}_{1\theta}
-\frac{\eta^3}{R^4} p^{b}_{0\theta}  \nonumber \\
&&  + \, \nu\Biggl[-\frac{1}{R}v^{b}_{2\eta}-\frac{\eta}{R^2} v^{b}_{1\eta}
-\frac{\eta^2}{R^3} v^{b}_{0\eta} + \frac{1}{R^2}\left(v^{b}_{1\theta\theta}-v^{b}_{1}\right)
+2\frac{\eta}{R^3}\left(v^{b}_{0\theta\theta}-v^{b}_{0}\right)+\frac{2}{R^2}u^{b}_{0\theta}\Biggr], \quad\quad  \label{A12} \\
G^{b}_{1}&=&0, \label{A13} \\
G^{b}_{2}&=&u^{b}_{0\tau}-\nu u^{b}_{0\eta\eta}, \label{A14} \\
G^{b}_{3}&=&u^{b}_{1\tau}-\nu u^{b}_{1\eta\eta}-
u^{i}_{0}\!\!\bigm\vert_{r=R}u^{b}_{0\eta}+
u^{i}_{0\theta}\!\!\bigm\vert_{r=R}v^{b}_{0}  \nonumber \\
&&-\frac{1}{R}\left((v^{b}_{0})^2+ 2v^{i}_{0}\!\!\bigm\vert_{r=R}v^{b}_{0}\right)
+\nu\left(\frac{1}{R}u^{b}_{0\eta}+\frac{2}{R^2}v^{b}_{0\theta}\right)  , \label{A15} \\
H^{b}_{1} &=&-\frac{\eta}{R^2}v^{b}_{0\theta}-\frac{1}{R}u^{b}_{0}, \label{A16} \\
H^{b}_{2}&=&-\frac{\eta}{R^2}v^{b}_{1\theta}-\frac{\eta^2}{R^3}v^{b}_{0\theta}-\frac{1}{R}u^{b}_{1}
-\frac{\eta}{R^2}u^{b}_{0} , \label{A17} \\
H^{b}_{3}&=&-\frac{\eta}{R^2}v^{b}_{2\theta}-\frac{\eta^2}{R^3}v^{b}_{1\theta}
-\frac{\eta^3}{R^4}v^{b}_{0\theta}-\frac{1}{R}u^{b}_{2}
-\frac{\eta}{R^2}u^{b}_{1}-\frac{\eta^2}{R^3}u^{b}_{0} . \label{A18}
\end{eqnarray}

\vskip 3mm
\noindent
Functions $Q^{b}_{k}$ and $S^{b}_{k}$ in Eqs. (\ref{3.40}) and (\ref{3.41})
are given by
\begin{eqnarray}
Q_{1} &=&0 ,  \label{A19} \\
Q_{2} &=&-\cos\theta \, f\left(u^{i}_{0r}\!\!\bigm\vert_{r=1}
+ u^{a}_{0\xi}\!\!\bigm\vert_{\xi=0}\right)+\sin\theta \, f \, u^{i}_{0\theta}\!\!\bigm\vert_{r=1} \label{A20} \\
Q_{3} &=&-\cos\theta \, f\left(u^{i}_{1r}\!\!\bigm\vert_{r=1}
+ u^{a}_{1\xi}\!\!\bigm\vert_{\xi=0}\right)
+\sin\theta \, f\left(u^{i}_{1\theta}\!\!\bigm\vert_{r=1}+
u^{a}_{0\theta}\!\!\bigm\vert_{\xi=0}\right) \nonumber \\
&&-\sin\theta \, f\left(v^{i}_{1}\!\!\bigm\vert_{r=1}+ v^{a}_{1}\!\!\bigm\vert_{\xi=0}\right)
- \cos^2\theta \frac{f^2}{2}u^{a}_{0\xi\xi}\!\!\bigm\vert_{\xi=0}
-\sin\theta\cos\theta \, f^2v^{a}_{0\xi}\!\!\bigm\vert_{\xi=0} , \label{A21} \\
S_{1} &=&- \cos\theta \, f(\tau) v^{a}_{0\xi}\!\!\bigm\vert_{\xi=0}, \label{A22} \\
S_{2} &=&- \cos\theta \, f\left( v^{i}_{0r}\!\!\bigm\vert_{r=1}
+v^{a}_{1\xi}\!\!\bigm\vert_{\xi=0}\right)
+ \sin\theta \, f\left( v^{i}_{0\theta}\!\!\bigm\vert_{r=1}
+v^{a}_{0\theta}\!\!\bigm\vert_{\xi=0}\right) \nonumber \\
&&+\sin\theta \, f \, u^{i}_{0}\!\!\bigm\vert_{r=1}
- \cos^2\theta \frac{f^2}{2}v^{a}_{0\xi\xi}\!\!\bigm\vert_{\xi=0} , \label{A23} \\
S_{3} &=&- \cos\theta \, f\left( v^{i}_{1r}\!\!\bigm\vert_{r=1}
+v^{a}_{2\xi}\!\!\bigm\vert_{\xi=0}\right)
+ \sin\theta \, f\left( v^{i}_{1\theta}\!\!\bigm\vert_{r=1}
+v^{a}_{1\theta}\!\!\bigm\vert_{\xi=0}\right) \nonumber \\
&&+\sin\theta \, f \left(u^{i}_{1}\!\!\bigm\vert_{r=1}+u^{a}_{0}\!\!\bigm\vert_{\xi=0}\right)
- \cos^2\theta \frac{f^2}{2}v^{a}_{1\xi\xi}\!\!\bigm\vert_{\xi=0} \nonumber \\
&&+\sin\theta\cos\theta \, f^2 v^{a}_{0\xi\theta}\!\!\bigm\vert_{\xi=0}
-\sin^2\theta \frac{f^2}{2}v^{a}_{0\xi}\!\!\bigm\vert_{\xi=0}
- \cos^3\theta \frac{f^3}{6}v^{a}_{0\xi\xi\xi}\!\!\bigm\vert_{\xi=0} . \quad\quad \label{A24}
\end{eqnarray}
Explicit formulae for constants $C_{1}$, $C_{2}$, $C_{3}$ and $C_{4}$ in Eq. (\ref{4.50})
are
\begin{eqnarray}
&&C_1 = R^6(R^2+2)+2R^2+1, \quad
C_2 = -R^8-R^2,  \nonumber \\
&&C_3 = -R^4(2R^2+1)-R^2-2 ,  \quad
C_4 = R^4+1. \label{A25}
\end{eqnarray}
Constants $D_{1}$, $D_{2}$, $D_{3}$ and $D_{4}$ in Eq. (\ref{4.62})
are given by
\begin{eqnarray}
D_1 &=& -(75R^{10}-75R^9+138R^8-158R^7 \nonumber \\
&& + \, 106R^6-134R^5+131R^4-79R^3+80R^2-40R+40), \nonumber \\
D_2 &=& R^2(53R^8-53R^7+32R^6-44R^5+39R^4-51R^3+52R^2-14R+14), \nonumber \\
D_3 &=& 2(75R^8-77R^7+58R^6-74R^5+59R^4-52R^3+53R^2-40R+40), \nonumber \\
D_4 &=& - (75R^6-79R^5+31R^4-43R^3+44R^2-40R+40). \label{A26}
\end{eqnarray}


\setcounter{equation}{0}
\renewcommand{\theequation}{B.\arabic{equation}}

\section{Appendix B. Calculation of the averaged Lagrangian velocity}


\noindent
In polar coordinates, Eq. (\ref{4.65}) is equivalent to
\begin{eqnarray}
&&r_{\tau}=\epsilon^2\left[u(r,\theta,\tau)-r_{t}\right]. \label{5.2} \\
&&r\theta_{\tau}=\epsilon^2\left[v(r,\theta,\tau)-r\theta_{t}\right]. \label{5.3}
\end{eqnarray}
We already know that
\begin{eqnarray}
&&u=u^{i}_{0} + \eps\left(u^{i}_{1}+
u^{a}_{0}+ u^{b}_{0}\right)+\eps^{2}\left(u^{i}_{2}+
u^{a}_{1}+ u^{b}_{1}\right)+\dots, \label{5.4} \\
&&v=v^{i}_{0}+v^{a}_{0}+ v^{b}_{0}
+ \eps\left(v^{i}_{1}+
v^{a}_{1}+ v^{b}_{1}\right)+\eps^{2}\left(v^{i}_{2}+
v^{a}_{2}+ v^{b}_{2}\right)+\dots \label{5.5}
\end{eqnarray}
We seek the solution of (\ref{5.2}) and (\ref{5.3}) in the form
\begin{equation}
r=r_{0}+\eps r_{1}+\eps^2 r_{2}+\dots, \quad
\theta=\theta_{0}+\eps \theta_{1}+\eps^2 \theta_{2}+\dots \label{5.6}
\end{equation}
On substituting (\ref{5.4})--(\ref{5.6}) in (\ref{5.2}) and (\ref{5.3}) and
collecting terms of equal powers in $\eps$, we obtain the following sequence of equations:
\begin{eqnarray}
&&r_{0\tau}=0, \quad r_{0}\theta_{0\tau}=0, \label{5.7} \\
&&r_{1\tau}=0, \quad
r_{0}\theta_{1\tau}+r_{1}\theta_{0\tau}=0, \label{5.8}
\end{eqnarray}
and
\begin{eqnarray}
&&\pr_{\tau}r_{n}+\pr_{t}r_{n-2}=U_{n-2}, \label{5.13} \\
&&\sum_{m=0}^{n}r_{m}\pr_{\tau}\theta_{n-m}+
\sum_{m=0}^{n-2}r_{m}\pr_{t}\theta_{n-m-2}=V_{n-2}, \label{5.14}
\end{eqnarray}
for $n=2,3,\dots$
Functions $U_{n}$ and $V_{n}$ are obtained by substitution of
(\ref{5.4})--(\ref{5.6}) in the right sides of Eqs. (\ref{5.2}) and (\ref{5.3}) and  subsequent Taylor's expansion
of all terms. There is a subtle technical trick here. Functions which depend on boundary layer
variables $\xi=(r-1)/\eps$ and $\eta=(R-r)/\eps$ are not analytic at $\eps=0$. Therefore, for these functions
we use Taylor's expansions
at $r=r_{0}+\eps r_{1}$ rather than at $r=r_{0}+\eps r_{1}$. For example,
\begin{eqnarray}
u_{0}^{a}(\xi,\theta,\tau) &=& u_{0}^{a}\left(\frac{r_{0}-1}{\eps}+r_{1}+\eps r_{2}+
\dots, \theta_{0}+\eps \theta_{1}+\dots,\tau\right) \nonumber \\
&=&
u_{0}^{a}(\xi^{*},\theta_{0},\tau)+
\eps \left[r_{2} \, u_{0\xi}^{a}(\xi^{*},\theta_{0},\tau)+ \theta_{1} \,
u_{0\theta}^{a}(\xi^{*},\theta_{0},\tau)\right] +\dots \nonumber
\end{eqnarray}
where $\xi^{*}=\frac{r_{0}-1}{\eps}+r_{1}$. Below we use the following notation:
$(\cdot)\bigm\vert_{0}=(\cdot)\bigm\vert_{r=r_{0}, \theta=\theta_{0}}$
and
\[
(\cdot)\bigm\vert_{*}=(\cdot)\bigm\vert_{\xi=\frac{r_{0}-1}{\eps}+r_{1}, \, \theta=\theta_{0}}
\quad {\rm or}\quad
(\cdot)\bigm\vert_{*}=(\cdot)\bigm\vert_{\eta=\frac{R-r_{0}}{\eps}-r_{1}, \, \theta=\theta_{0}}
\]
depending on whether the quantity being evaluated is a function of $\xi$ or $\eta$.
The explicit formulae for $U_{n}$ and $V_{n}$ are
\begin{eqnarray}
U_{0}&=& u_{0}^{i}\!\bigm\vert_{0}, \quad
V_{0}=v_{0}^{i}\!\bigm\vert_{0} + v_{0}^{a}\!\bigm\vert_{*}+v_{0}^{b}\!\bigm\vert_{*}, \label{5.15} \\
U_{1}&=& \left[u_{1}^{i}+
r_{1}u_{0r}^{i}+\theta_{1}u_{0\theta}^{i}\right]\!\bigm\vert_{0}
+ u_{0}^{a}\!\bigm\vert_{*}+u_{0}^{b}\!\bigm\vert_{*}, \label{5.16} \\
V_{1}&=& \left[v_{1}^{i}+
r_{1}v_{0r}^{i}+\theta_{1}v_{0\theta}^{i}\right]\!\bigm\vert_{0}
+ \left[v_{1}^{a}+
r_{2}v_{0\xi}^{a}+\theta_{1}v_{0\theta}^{a}\right]\!\bigm\vert_{*}
+ \left[v_{1}^{b}-
r_{2}v_{0\eta}^{b}+\theta_{1}v_{0\theta}^{b}\right]\!\Bigm\vert_{*}, \label{5.17} \\
U_{2}&=& \left[u_{2}^{i}+
r_{1}u_{1r}^{i}+\theta_{1}u_{1\theta}^{i}
+\frac{1}{2}\left(r_{1}^2u_{0rr}^{i}+2r_{1}\theta_{1}u_{0r\theta}^{i}
+\theta_{1}^2u_{0\theta\theta}^{i}\right)+r_{2}u_{0r}^{i}+\theta_{2}u_{0\theta}^{i}
\right]\!\biggm\vert_{0} \nonumber \\
&&+\left[u_{1}^{a}+
r_{2}u_{0\xi}^{a}+\theta_{1}u_{0\theta}^{a}\right]\!\bigm\vert_{*}
+ \left[u_{1}^{b}-
r_{2}u_{0\eta}^{b}+\theta_{1}u_{0\theta}^{b}\right]\!\Bigm\vert_{*}, \label{5.18} \\
V_{2}&=& \left[v_{2}^{i}+
r_{1}v_{1r}^{i}+\theta_{1}v_{1\theta}^{i}
+\frac{1}{2}\left(r_{1}^2v_{0rr}^{i}+2r_{1}\theta_{1}v_{0r\theta}^{i}
+\theta_{1}^2v_{0\theta\theta}^{i}\right)+r_{2}v_{0r}^{i}+\theta_{2}v_{0\theta}^{i}
\right]\!\biggm\vert_{0} \nonumber \\
&&+ \left[v_{2}^{a}+
r_{2}v_{1\xi}^{a}+\theta_{1}v_{1\theta}^{a}
\frac{1}{2}\left(r_{2}^2v_{0\xi\xi}^{a}+2r_{2}\theta_{1}v_{0\xi\theta}^{a}
+\theta_{1}^2v_{0\theta\theta}^{a}\right)
+r_{3}v_{0\xi}^{a}+\theta_{2}v_{0\theta}^{a}
\right]\!\biggm\vert_{*} \nonumber \\
&&+\left[v_{2}^{b}-
r_{2}v_{1\eta}^{b}+\theta_{1}v_{1\theta}^{b}+
\frac{1}{2}\left(r_{2}^2v_{0\eta\eta}^{b}-2r_{2}\theta_{1}v_{0\eta\theta}^{b}
+\theta_{1}^2v_{0\theta\theta}^{b}\right)
-r_{3}v_{0\eta}^{b}+\theta_{2}v_{0\theta}^{b}
\right]\!\biggm\vert_{*}. \quad\quad \label{5.19}
\end{eqnarray}
It follows from Eqs. (\ref{5.7}) and (\ref{5.8}) that
$r_{0}$, $\theta_{0}$, $r_{1}$ and $\theta_{1}$ do not depend on fast time $\tau$, i.e.
$r_{0}=r_{0}(t)$, $\theta_{0}=\theta_{0}(t)$, $r_{1}=r_{1}(t)$ and $\theta_{1}=\theta_{1}(t)$.
Equations (\ref{5.15}) and the fact that
$\bar{u}_{0}^{i}=0$, $\bar{v}_{0}^{i}=0$, $\bar{v}_{0}^{a}=0$, $\bar{v}_{0}^{b}=0$ imply that
$\bar{U}_{0}=0$ and $\bar{V}_{0}=0$. Then, averaging Eqs. (\ref{5.13}) and (\ref{5.14}) (for $n=2$),
we find that
$r_{0t}=0$ and $r_{0}\theta_{0t}=0$. Therefore, $r_{0}={\rm const}$ and $\theta_{0}={\rm const}$,
and we use $r_{0}$ and $\theta_{0}$ to identify fluid particles:
$(r(r_{0},\theta_{0},t,\tau), \theta(r_{0},\theta_{0},t,\tau))$
represents the trajectory of a fluid particle whose averaged (in $\tau$) position at $t=0$ was
$(r_{0},\theta_{0})$.
From Eq. (\ref{5.16}), we deduce that
$\bar{U}_{1}=0$. Averaging Eq. (\ref{5.13}) for $n=3$, we find that
$r_{1t}=0$,
so that $r_{1}={\rm const}$. We choose
$r_{1}=0$.
Note that this implies that $(\cdot)\bigm\vert_{*}=(\cdot)\bigm\vert_{0}$ in Eqs. (\ref{5.17})--(\ref{5.19}).
Averaging Eq. (\ref{5.14}) for $n=3$ yields
\begin{equation}
r_{0}\theta_{1t}=\left[\bar{v}_{1}^{a}+
\overline{r_{2}v_{0\xi}^{a}}\right]\!\Bigm\vert_{0}
- \, \overline{r_{2}v_{0\eta}^{b}}\Bigm\vert_{0}. \label{5.23}
\end{equation}
To proceed further, we need to compute $\overline{r_{2}v_{0\xi}^{a}}\Bigm\vert_{0}$ and
$\overline{r_{2}v_{0\eta}^{b}}\Bigm\vert_{0}$. First we observe that the averaged part of $r_{2}$
does not make any contribution to these two terms, i.e.
$\overline{r_{2}v_{0\xi}^{a}}\Bigm\vert_{0}=\overline{\tilde{r}_{2}v_{0\xi}^{a}}\Bigm\vert_{0}$,
$\overline{r_{2}v_{0\eta}^{b}}\Bigm\vert_{0}=\overline{\tilde{r}_{2}v_{0\eta}^{b}}\Bigm\vert_{0}$.
The oscillatory part of $r_{2}$ satisfies the equation
\[
\tilde{r}_{2\tau}=U_{0}=u_{0}^{i}\!\bigm\vert_{0}=u_{0}^{i}(r_{0},\theta_{0},\tau)
\]
(which is simply the oscillatory part of Eq. (\ref{5.13}) for $n=2$).
Substitution of (\ref{4.1}) yields
\begin{equation}
\tilde{r}_{2}=\frac{f(\tau)}{R^2-1}\left(\frac{R^2}{r_{0}^2}-1\right)\cos\theta_{0}. \label{5.24}
\end{equation}
This implies that
\[
\overline{r_{2}v_{0\xi}^{a}}\Bigm\vert_{0}=\frac{\cos\theta_{0}}{R^2-1}\left(\frac{R^2}{r_{0}^2}-1\right)
\overline{f \, v_{0\xi}^{a}}, \quad\quad
\overline{r_{2}v_{0\eta}^{b}}\Bigm\vert_{0}=\frac{\cos\theta_{0}}{R^2-1}\left(\frac{R^2}{r_{0}^2}-1\right)
\overline{f \, v_{0\eta}^{b}}.
\]
On the right sides of these equations we have functions $v_{0\xi}^{a}$ and $v_{0\eta}^{b}$ which are nonzero
only within boundary layers with thickness $O(\eps)$. It is therefore natural (and consistent with our
asymptotic expansion for the velocity) to replace $r_{0}$ by $1+\eps\xi_{0}$ in the first equation
and by $R-\eps\eta_{0}$ in the second equation (where $\xi_{0}=(r_{0}-1)/\eps$ and
$\eta_{0}=(R-r_{0})/\eps$) and expand everything in Taylor's series. This leads to the expressions
\begin{eqnarray}
\overline{r_{2}v_{0\xi}^{a}}\Bigm\vert_{0} &=& \cos\theta_{0}\overline{f \, v_{0\xi}^{a}}
\left(1-\eps\frac{2\xi_{0}R^2}{R^2-1} +O(\eps^2)\right), \label{5.25} \\
\overline{r_{2}v_{0\eta}^{b}}\Bigm\vert_{0} &=& \eps\frac{2\eta_{0}}{R(R^2-1)}
\cos\theta_{0}\overline{f \, v_{0\eta}^{b}} +O(\eps^2). \label{5.26}
\end{eqnarray}
Now we substitute $O(1)$ terms from (\ref{5.25}) and (\ref{5.26}) in Eq. (\ref{5.23})
and move higher order terms to appropriate higher
order equations. Equation (\ref{5.23}) becomes
\[
r_{0}\theta_{1t}=\left[\bar{v}_{1}^{a}+
\cos_{0}\overline{f \, v_{0\xi}^{a}}\right]\!\Bigm\vert_{0}.
\]
The first term on the right side of this equation represents the averaged Eulerian velocity, while
the second term is associated with the Stokes drift.
Finally, substitution of (\ref{4.20}) reduces this equation to $r_{0}\theta_{1t}=0$,
which, in turn, implies that $\theta_{1}={\rm const}$. Again, we choose
$\theta_{1}=0$.
Thus, the averaged Lagrangian velocity has no $O(\eps)$ boundary layer term.
Averaging the equation for $r_{4}$ yields
\begin{equation}
\bar{r}_{2t}=\left[\bar{u}_{2}^{i}+
\overline{\tilde{r}_{2}u_{0r}^{i}}+
\overline{\tilde{\theta}_{2}u_{0\theta}^{i}}\right]\!\Bigm\vert_{0}
+\left[\bar{u}_{1}^{a}+
\overline{\tilde{r}_{2}u_{0\xi}^{a}}\right]\!\Bigm\vert_{0}
- \overline{\tilde{r}_{2}u_{0\eta}^{b}}\!\Bigm\vert_{0}. \label{5.28}
\end{equation}
Now we need an explicit formula for $\tilde{\theta}_{2}$. The equation for $\tilde{\theta}_{2}$
is
\[
\tilde{\theta}_{2\tau}=\frac{1}{r_{0}}\left(v_{0}^{i}\!\bigm\vert_{0} + v_{0}^{a}\!\bigm\vert_{0}
+v_{0}^{b}\!\bigm\vert_{0}\right).
\]
On substituting Eqs. (\ref{4.1}), (\ref{4.4}), (\ref{4.11}) and integrating over $\tau$, we obtain
\begin{eqnarray}
\tilde{\theta}_{2} &=&\tilde{\theta}_{2}^{i}+\tilde{\theta}_{2}^{a}+\tilde{\theta}_{2}^{b} \label{5.29} \\
\tilde{\theta}_{2}^{i} &=&\frac{\sin\theta_{0}}{R^2-1}\frac{1}{r_{0}}
\left(\frac{R^2}{r_{0}^2}+1\right)f(\tau), \label{5.30} \\
\tilde{\theta}_{2}^{a} &=&-\frac{2R^2\sin\theta_{0}}{R^2-1}
Re\left(Ce^{-\gamma\xi_{0}+i\tau}\right) + O(\eps), \label{5.31} \\
\tilde{\theta}_{2}^{b} &=&-\frac{2\sin\theta_{0}}{R(R^2-1)}
Re\left(Ce^{-\gamma\eta_{0}+i\tau}\right) + O(\eps). \label{5.32}
\end{eqnarray}
Now we return to Eq. (\ref{5.28}). It follows from (\ref{4.1}), (\ref{4.6}), (\ref{4.12}), (\ref{5.24}),
(\ref{5.30})--(\ref{5.32}) that
\begin{eqnarray}
&&\overline{\tilde{r}_{2}u_{0r}^{i}}\Bigm\vert_{0} =0, \quad
\overline{\tilde{\theta}_{2}^{i}u_{0\theta}^{i}}\Bigm\vert_{0}
=0, \quad \overline{\tilde{\theta}_{2}^{b}u_{0\theta}^{i}}\Bigm\vert_{0} =O(\eps),
\quad \overline{\tilde{r}_{2}u_{0\eta}^{b}}\Bigm\vert_{0}=O(\eps), \quad \quad \label{5.35} \\
&&\overline{\tilde{\theta}_{2}^{a}u_{0\theta}^{i}}\Bigm\vert_{0} =
\sin\theta_{0} \overline{f \, v_{0}^{a}}+O(\eps),
\quad
\overline{\tilde{r}_{2}u_{0\xi}^{a}}\Bigm\vert_{0}=-\cos\theta_{0}\overline{f \, v_{0\theta}^{a}}+O(\eps). \label{5.36}
\end{eqnarray}
After substitution of (\ref{5.35}) and (\ref{5.36}) in Eq. (\ref{5.28}), it reduces to
\begin{equation}
\bar{r}_{2t}=\bar{u}_{2}^{i}(r_{0},\theta_{0}). \label{5.39}
\end{equation}
Note that the right side of this equation is different from what we
would have if we used the averaged Eulerian velocity: the second order term in the expansion of $\bar{u}$
is $\bar{u}_{2}^{i}+\bar{u}_{1}^{b}$ rather than $\bar{u}_{2}^{i}$. This means that the Stokes drift
kills the boundary layer term.

\vskip 3mm
\noindent
Now let us deduce equation for $\bar{\theta}_{2}$.
On averaging the equation for $\theta_{4}$ and using (\ref{5.19}),
we obtain
\begin{eqnarray}
r_{0}\bar{\theta}_{2t} &=&
\left[\bar{v}_{2}^{i}+
\overline{{r}_{2}v_{0r}^{i}}+
\overline{{\theta}_{2}v_{0\theta}^{i}}\right]
+\left[\bar{v}_{2}^{a}+
\overline{{r}_{2}v_{1\xi}^{a}}+\frac{1}{2}\overline{{r}_{2}^2 v_{0\xi\xi}^{a}}
+\overline{{r}_{3}v_{0\xi}^{a}}+\overline{{\theta}_{2}v_{0\xi}^{a}}\right] \nonumber \\
&&+\left[\bar{v}_{2}^{b}-
\overline{{r}_{2}v_{1\eta}^{b}}+\frac{1}{2}\overline{{r}_{2}^2 v_{0\eta\eta}^{b}}
-\overline{{r}_{3}v_{0\eta}^{b}}+\overline{{\theta}_{2}v_{0\eta}^{b}}\right]
-\overline{r_2\theta_{2\tau}}
 + \, T_{2}, \label{5.40}
\end{eqnarray}
where all quantities on the right side of this equation are evaluated at $(r_{0},\theta_{0})$
and where
\begin{equation}
T_{2}=-\cos\theta_{0}\frac{2\xi_{0}R^2}{R^2-1}\overline{f \, v_{0\xi}^{a}}
+\cos\theta_{0}\frac{2\eta_{0}}{R(R^2-1)}
\overline{f \, v_{0\eta}^{b}} \label{5.41}
\end{equation}
represent the contribution that comes from $O(\eps)$ terms in
(\ref{5.25}) and (\ref{5.26}).
To proceed further, we need an explicit formula for $\tilde{r}_{3}$.
We have the equation
$\tilde{r}_{3\tau}=u_{1}^{i}\!\bigm\vert_{0} + u_{0}^{a}\!\bigm\vert_{0}
+u_{0}^{b}\!\bigm\vert_{0}$, from which,
on substituting (\ref{4.6}), (\ref{4.12}), (\ref{4.16}) and integrating over $\tau$, we obtain
\begin{eqnarray}
\tilde{r}_{3} &=& \tilde{r}_{3}^{i}+\tilde{r}_{3}^{a}+\tilde{r}_{3}^{b}, \label{5.42} \\
\tilde{r}_{3}^{i} &=&-\frac{2R}{(R^2-1)^2}
\left[R+1 - \frac{R^3+1}{r^2}\right]Re
\left(\frac{C}{\gamma}e^{i\tau}\right)\cos\theta_{0}, \label{5.43} \\
\tilde{r}_{3}^{a} &=&-\frac{2R^2}{R^2-1}
Re\left(\frac{C}{\gamma}e^{-\gamma\xi_{0}+i\tau}\right)\cos\theta_{0} , \label{5.44} \\
\tilde{r}_{3}^{b} &=&\frac{2}{R(R^2-1)}
Re\left(\frac{C}{\gamma}e^{-\gamma\eta_{0}+i\tau}\right)\cos\theta_{0}. \label{5.45}
\end{eqnarray}
After tedious but elementary calculations with the help of (\ref{5.24}), (\ref{5.29})--(\ref{5.32}),
(\ref{5.42})--(\ref{5.45}), Eq. (\ref{5.40}) can be reduced to
\[
r_{0}\bar{\theta}_{2t} =
\bar{v}_{2}^{i}+\bar{v}_{2}^{a}+\bar{v}_{2}^{b}+\bar{v}_{2}^{s}
\]
where $\bar{v}_{2}^{s}$ represents the Stokes drift velocity of the fluid particles and is given by
\begin{eqnarray}
\bar{v}_{2}^{s} &=&
\cos\theta_{0}\left[\overline{f \, v_{1\xi}^{a}}-\overline{f \, v_{0}^{a}}\right]
+  \frac{2\cos\theta_{0}}{R^2-1}\left\{R^2
\left[\overline{g \, v_{0\xi}^{a}} - \xi_{0} \, \overline{f \, v_{0\xi}^{a}}
-\overline{Re\left(\frac{C}{\gamma}e^{-\gamma\xi_{0}+i\tau}\right) \, v_{0\xi}^{a}}
\right]\right. \quad\quad \nonumber \\
&&+
\left.\frac{1}{R}\left[\overline{g \, v_{0\eta}^{b}} - \eta_{0} \, \overline{f \, v_{0\eta}^{b}}
-\overline{Re\left(\frac{C}{\gamma}e^{-\gamma\eta_{0}+i\tau}\right) \, v_{0\eta}^{b}}\right]
\right\}. \label{5.46}
\end{eqnarray}
Here $g(\tau)=Re(C e^{i\tau}/\gamma)$.
In order to obtain the $O(\eps^3)$ term in the expansion for the stream function,
we need to compute the components of the averaged third-order Lagrangian velocity in the outer
flow.
Averaging the equations for $r_{5}$ and $\theta_{5}$ and ignoring all boundary layer terms, we obtain
\begin{eqnarray}
\bar{r}_{3t}^{i}&=&\bar{u}_{3}^{i}+
\overline{r_{2}^{i}u_{1r}^{i}}+\overline{r_{3}^{i}u_{0r}^{i}}+
\overline{\theta_{2}^{i}u_{1\theta}^{i}} +\overline{\theta_{3}^{i}u_{0\theta}^{i}}, \label{5.55} \\
r_{0}\bar{\theta}_{3t}^{i}&=&-\overline{r_{2}^{i}\theta_{3\tau}^{i}}-\overline{r_{3}^{i}\theta_{2\tau}^{i}}
+\bar{v}_{3}^{i}+
\overline{r_{2}^{i}v_{1r}^{i}}+\overline{r_{3}^{i}v_{0r}^{i}}+
\overline{\theta_{2}^{i}v_{1\theta}^{i}} +\overline{\theta_{3}^{i}v_{0\theta}^{i}}. \label{5.56}
\end{eqnarray}
All terms on the right sides of these equations are evaluated at $(r_{0}, \theta_{0})$.
Now we need to find $\tilde{\theta}_{3}^{i}$. The equation for $\tilde{\theta}_{3}^{i}$ is obtained
from (\ref{5.14}) for $n=3$ by taking its oscillatory part and ignoring all boundary layer terms:
\[
r_{0}\tilde{\theta}_{3\tau}^{i}=\tilde{v}_{1}^{i}(r_{0},\theta_{0},\tau) .
\]
Inserting (4.17) into this equation and integrating over $\tau$, we find that
\begin{equation}
\tilde{\theta}_{3}^{i}= \frac{2R}{(R^2-1)^2}\left((R+1) +\frac{R^3+1}{r^2}\right)g(\tau) \,
\sin\theta . \label{5.57}
\end{equation}
Substitution of (\ref{4.1}), (\ref{4.2}), (\ref{4.19}), (\ref{4.20}), (\ref{5.24}), (\ref{5.30}),
(\ref{5.43}) and (\ref{5.57}) in
Eqs. (\ref{5.55}) and (\ref{5.56}) yields
\begin{eqnarray}
\bar{r}_{3t}^{i}&=&\bar{u}_{3}^{i}+
\frac{4R}{(R^2-1)^2} \, \overline{f^{\prime} \, g} \, \frac{\cos 2\theta}{r^3}, \label{5.58} \\
r_{0}\bar{\theta}_{3t}^{i}&=&
\bar{v}_{3}^{i}+
\frac{4R}{(R^2-1)^2} \, \overline{f^{\prime} \, g} \, \frac{\sin 2\theta}{r^3}. \label{5.59}
\end{eqnarray}
Further calculations yield
\begin{eqnarray}
\left(\bar{u}_{3}^{i}\right)^{L}&=& \, \frac{\sqrt{2\nu}}{4R(R-1)(R^2-1)^4}
\left(\frac{D_1}{r}+\frac{\hat{D}_2}{r^3}+D_3 r+D_4 r^3\right)
\cos 2\theta, \label{5.60} \\
\left(\bar{v}_{3}^{i}\right)^{L}&=& \, \frac{\sqrt{2\nu}}{4R(R-1)(R^2-1)^4}
\left(\frac{\hat{D}_2}{r^3}-D_3 r-2D_4 r^3\right)\sin 2\theta,  \label{5.61}
\end{eqnarray}
where
\begin{equation}
\hat{D}_2=D_2-4R^2(R-1)(R^2-1)^2. \label{BBB}
\end{equation}
Hence,
\begin{equation}
\left(\bar{\psi}^{i}_{3}\right)^{L}=\frac{\sqrt{2\nu}}{8R(R-1)(R^2-1)^4}
\left(D_1+\frac{\hat{D}_2}{r^2}+D_3 r^2+D_4 r^4\right)\sin 2\theta \label{5.62}
\end{equation}

\end{document}